\documentclass[fleqn,usenatbib]{mnras}
\usepackage{newtxtext,newtxmath}
\usepackage[T1]{fontenc}

\usepackage{graphicx}
\usepackage{amsmath}
\usepackage{comment}

\title[Dark matter back-reaction in CAMELS]{Cosmological back-reaction of baryons on dark matter in the CAMELS simulations}

\author[Gebhardt et al.]{Matthew Gebhardt,$^{1}$\thanks{E-mail: mattgebhardt2@gmail.com}
Daniel Anglés-Alcázar$^{1}$,
Shy Genel$^{2,3}$,
Daisuke Nagai$^{4,5}$,
Boon Kiat Oh$^{1}$,
\newauthor 
Isabel Medlock$^{5}$,
Jonathan Mercedes-Feliz$^{1}$,
Sagan Sutherland$^{1}$,
Max E. Lee$^{3}$,
Xavier Sims$^{1}$,
\newauthor 
Christopher C. Lovell$^{6}$,
David N. Spergel$^{2,7}$,
Romeel Davé$^{8,9,10}$,
Matthieu Schaller$^{11,12}$,
\newauthor 
Joop Schaye$^{12}$,
Francisco Villaescusa-Navarro$^{2,7}$
\\
$^{1}$Department of Physics, University of Connecticut, 196 Auditorium Road, U-3046, Storrs, CT 06269-3046, USA\\
$^{2}$Center for Computational Astrophysics, Flatiron Institute, 162 Fifth Avenue, New York, NY 10010, USA\\
$^{3}$Department of Astronomy, Columbia University, New York, NY 10027, USA\\
$^{4}$Department of Physics, Yale University, New Haven, CT 06520, USA\\
$^{5}$Department of Astronomy, Yale University, New Haven, CT 06520, USA\\
$^{6}$Kavli Institute for Cosmology, University of Cambridge, Cambridge CB3 0HA, UK\\
$^{7}$Department of Astrophysical Sciences, Princeton University, 4 Ivy Lane, Princeton, NJ 08544 USA\\
$^{8}$Institute for Astronomy, Royal Observatory, University of Edinburgh, Edinburgh EH9 3HJ, UK\\
$^{9}$University of the Western Cape, Department of Physics and Astronomy, Bellville, Cape Town 7535, South Africa\\
$^{10}$South African Astronomical Observatories, Observatory, Cape Town 7925, South Africa\\
$^{11}$Lorentz Institute for Theoretical Physics, Leiden University, PO Box 9506, NL-2300 RA Leiden, the Netherlands\\
$^{12}$Leiden Observatory, Leiden University, PO Box 9513, NL-2300 RA Leiden, the Netherlands\\
}

\date{Accepted XXX. Received YYY; in original form ZZZ}

\pubyear{2025}

\begin{document}
\label{firstpage}
\pagerange{\pageref{firstpage}--\pageref{lastpage}}
\maketitle

\begin{abstract}
Baryonic processes such as radiative cooling and feedback from massive stars and active galactic nuclei (AGN) directly redistribute baryons in the Universe but also indirectly redistribute dark matter due to changes in the gravitational potential.
In this work, we investigate this ``back-reaction'' of baryons on dark matter using thousands of cosmological hydrodynamic simulations from the Cosmology and Astrophysics with MachinE Learning Simulations (CAMELS) project, including parameter variations in the SIMBA, IllustrisTNG, ASTRID, and Swift-EAGLE galaxy formation models. 
Matching haloes to corresponding $N$-body (dark matter-only) simulations, we find that virial masses decrease owing to the ejection of baryons by feedback. 
Relative to \textit{N}-body simulations, halo profiles show an increased dark matter density in the center (due to radiative cooling) and a decrease in density farther out (due to feedback), with both effects being strongest in SIMBA ($\gtrsim$\,450\% increase at $r \lesssim 0.01\,R_{\rm vir}$). 
The clustering of dark matter strongly responds to changes in baryonic physics, with dark matter power spectra in some simulations from each model showing as much as 20\% suppression or increase in power at $k \sim 10\,h$\,Mpc$^{-1}$ relative to $N$-body simulations. We find that the dark matter back-reaction depends intrinsically on cosmology ($\Omega_{\rm m}$ and $\sigma_{8}$) at fixed baryonic physics, and varies strongly with the details of the feedback implementation. These results emphasize the need for marginalizing over uncertainties in baryonic physics to extract cosmological information from weak lensing surveys as well as their potential to constrain feedback models in galaxy evolution.
\end{abstract}

\begin{keywords}
galaxies:evolution -- galaxies:formation
\end{keywords}



\section{Introduction}

As cosmological surveys continue to improve, understanding the role that baryons play in the evolution of galaxies and the formation of structure in the Universe has become increasingly important. The hierarchal formation of structure in a purely cold dark matter (CDM) universe is well described by the Navarro-Frenk-White (NFW, \citealt{NFW_1997}) dark matter halo profile, which predicts high densities (a ``cusp'') in halo centers. In such a universe, the NFW profile is near-universal (applies to all haloes regardless of mass) due to the scale-free nature of gravity, resulting in self-similar substructures (``subhaloes'') within host haloes. The Einasto profile \citep{Einasto_1965} is another form of universal halo profile that has seen success in describing halo structure by including an additional parameter \citep[see e.g.][]{Merritt_2006_profiles}. 

These empirical profiles were derived from dark matter-only (hereafter ``$N$-body'') simulations \citep[see e.g.][]{Springel_2005c_nbody, Klypin_2011_nbody, Angulo_2012_nbody}, serving as a universal description of the structure of dark matter haloes collapsing under the influence of gravity. However, it is now well known that the structure of haloes can be quite different when baryons and their associated physics are included. Gas can cool and dissipate energy \citep{hoyle_1953}, which alters the structure at smaller scales and breaks the self-similarity found in $N$-body simulations \citep{white_and_rees_1978}. Furthermore, gas can be ejected to large distances due to feedback from massive stars \citep{Lynds_1963_SNobs, Madau_1996_IGM, Martin_1998_GW, Pettini_2001_GW} and accreting supermassive black holes (SMBHs, \citealt{Greene_2012_AGN, Cicone_2014_agnobs, Hlavacek-Larrondo2015, Garcia_2024_AGN, Vayner_2024_AGN}) in active galactic nuclei (AGN), key processes in the baryon cycle of galaxies that can significantly alter their evolution and the distribution of matter on larger scales. As a result, predictions from cosmological hydrodynamic simulations, which explicitly model gas (thermo)dynamics under the influence of gravity, hydrodynamic forces, and feedback, can differ from those of $N$-body simulations across a range of spatial scales, including the halo mass function (e.g. \citealt{Cui_2012_baryons_on_HMF, Cui_2014, Velliscig2014, Bocquet_2016, Schaye2023}), density profiles of individual haloes (e.g. \citealt{Duffy_2010_den_pro_BR, schaller_2015_eagle_analysis, Chan_2015_FIRE_BR, Sorini_2021_baryons, Sorini_2024_BR, Sorini_2024_simba}), and the total matter power spectrum (e.g. \citealt{Chisari_2019, van_Daalen_2020, Delgado_2023_powerspec, Pandey_2023_powerspec, Gebhardt_2024a, Bigwood2025_XFABLE, Schaller_2025_pk, Siegel2025, van_Daalen_2026}). 

Not only does baryonic physics alter the total matter distribution relative to $N$-body simulations by re-distributing baryons, but it also indirectly affects the dark matter distribution in a ``back-reaction'' effect. 
In the traditional adiabatic contraction framework, baryons can radiate away energy and deepen the gravitational potential in the centers of haloes, causing dark matter to contract inward \citep[see, e.g.,][]{Blumenthal1986,Gnedin_2004_contraction, sellwood_and_mcgaugh_2005_baryon_contract, Cardone_and_Sereno_2005_baryon_contract, klar_and_mucket_2008_baryon_contract}. In contrast, gas ejected to larger radial distances by feedback processes relaxes the gravitational potential farther out, causing dark matter to expand outward \citep{Mashchenko_2008_relax, Governato_2010_relax, Pontzen_2012_relax, Chan_2015_FIRE_BR}. These effects can play a significant role in observational constraints of the stellar initial mass function, the stellar mass-to-light ratio, and the dark matter content of galaxies \citep[e.g.,][]{Auger2010,Napolitano2010,Dutton2013,Tortora2014}.

It is crucial to understand the distribution of baryons and their back-reaction on dark matter for a number of additional reasons. First, weak lensing surveys such as DES \citep{Abbott_2022_DES}, KiDS \citep{Kuijken_2015_kids}, and Euclid \citep{Euclid_2022} observe the cosmic shear on galaxies due to intervening matter, mapping the projected total matter component. Ignoring the back-reaction effect can result in significant errors in the predictions of summary statistics such as the matter power spectrum \citep[see e.g.][]{van_Daalen_2011, van_Daalen_2020}, limiting the amount of information that can be extracted from cosmological surveys (\citealt{Semboloni2011, Chisari_2019}). Understanding the distribution of the dark matter component is also necessary for making accurate predictions for proposed dark matter detection experiments, which attempt to measure dark matter densities in the inner regions of haloes \citep[see e.g.][]{Stoehr_2003_DM_detection, Springel_2008_DM_detection, Grand_and_White_2021_DM_detection}. 

Unfortunately, much of the effect baryons have on the matter distribution depends on poorly-understood feedback processes. Implementing realistic feedback from massive stars and black holes in large-volume cosmological hydrodynamic simulations remains a major challenge due to the vast range of scales involved. Cosmological ``zoom-in'' simulations with increasing dynamic range (\citealt{DAA_2021_zoom, Hopkins_2023_fire3, Hopkins_2024_ForgedInFire}) can model more explicitly baryonic physics down to smaller scales, but cosmological simulations still rely on uncertain subgrid models to parametrize key baryonic processes (see \citealt{Somerville_2015, Crain_2023_SimReview} for reviews on cosmological hydrodynamic simulations). 

While typically calibrated to a common set of observables (such as the galaxy stellar mass function or the global star formation rate density at different redshifts), the results of cosmological hydrodynamic simulations can greatly vary from each other depending on baryonic physics implementation details. For example, \cite{Chisari_2019} explored the impact of baryons on the clustering of matter in the fiducial models of a variety of state-of-the-art cosmological hydrodynamic simulations and found that the suppression of the total matter power spectrum relative to $N$-body simulations can range from 10--30\% in different models at wave numbers from a few up to $20\,h\,\rm{Mpc^{-1}}$. In \cite{Gebhardt_2024a}, we investigated the large-scale spreading of baryons relative to dark matter in a variety of simulations and we found that while the SIMBA galaxy formation model (\citealt{simba_Dave_2019}) can spread $\sim$40\% of gas more than $1 \,{\rm Mpc}$ away (see also \citealt{Borrow_2020}), other models such as IllustrisTNG (\citealt{tng_Nelson_2018, Pillepich_2018_TNGpaper}) or ASTRID (\citealt{ni_2022_astridSMBH, bird_2022_astridgalaxys}) only eject $\sim$10\% of gas farther than $1 \,{\rm Mpc}$. Similarly, the closure radius of haloes, defined as the distance from a halo in which the enclosed baryon fraction equals the mean cosmic baryon fraction, differs significantly among simulations (\citealt{Ayromlou_2022_closureradius}). These results emphasize that feedback from supernovae (SNe) and AGN play an important role in shaping the distribution of matter in the Universe, but their effects can vary widely depending on the subgrid model implementation.

Previous works have investigated the back-reaction of baryons on dark matter using cosmological hydrodynamic simulations. \cite{Cui_2012_baryons_on_HMF}, found that halo masses are slightly increased in hydrodynamic simulations that include kinetic stellar feedback, but not AGN feedback. \cite{Cui_2014} found that including AGN feedback can significantly reduce the mass of haloes. Using simulations from the OWLS project \citep{Schaye_2010_owls}, \cite{Velliscig2014} found significant reduction in halo masses ($\sim20\%$) relative to corresponding $N$-body simulations, and noted that the removal of gas due to stellar and AGN feedback resulted in a corresponding expansion of dark matter in and around haloes. \cite{Sorini_2024_BR} showed that the masses of haloes in the IllustrisTNG and MilleniumTNG hydrodynamic simulations (both incorporating AGN feedback) were similarly reduced relative to their $N$-body counterparts, and the inclusion of baryons also resulted in steeper density profiles in the inner regions of haloes. \cite{Duffy_2010_den_pro_BR} used the wide physics variations in the OWLS suite of simulations to investigate the impact of baryons on dark matter halo profiles and found that higher halo baryon fractions generally correlated with more mass concentrated at small radii, with the back-reaction effect on dark matter depending strongly on the feedback model, particularly when including AGN feedback. Similarly, \cite{van_Daalen_2011} and later \cite{van_Daalen_2020} investigated the baryonic impact on the matter power spectrum. Using variations of the OWLS, cosmo-OWLS \citep{LeBrun2014}, and BAHAMAS \citep{McCarthy_2017_bahamas, McCarthy_2018_bahamas} simulations, they found that the dark matter power spectrum is significantly affected by AGN feedback, largely following the trends found for the total matter distribution. 

These previous results have emphasized the need for even larger suites of simulations varying assumptions about galaxy formation physics to quantify model uncertainties and extract cosmological information while marginalizing over baryonic effects. Addressing this need, the Cosmology and Astrophysics with MachinE Learning (CAMELS) project\footnote{\url{https://www.camel-simulations.org/}} \citep{Villaescusa_Navarro_2021, CAMELS_data_release} has recently produced the largest suite of state-of-the-art cosmological hydrodynamic simulations. Designed to train machine learning models for a variety of applications, CAMELS contains thousands of simulations varying initial conditions, subgrid physics implementations, and cosmological and astrophysical parameters. A key advantage of CAMELS is the ability to directly compare many different galaxy formation models at the same resolution and volume, all while varying their subgrid parameters. CAMELS has now been used for a variety of works including searching for optimal summary statistics for cosmological analysis \citep[e.g.][]{Nicola_2022_sumstat, Villaescusa-Navarro_2022_OneGalaxy, Lehman_2024_sum_stat}, cosmological parameter inference with machine learning techniques \citep[e.g.][]{Villaescusa-Navarro_2020_marginalization_example, Villaescusa-Navarro_2021_marginalization_fields2, Villaescusa-Navarro_2021_marginalization_fields, Villanueva-Domingo_2022_marginalizing_cosmicgraphs, Shao_2022_marginalization_subhaloes, Perez_2022_camelsSAM, deSanti_2023_fieldlevelgalaxies, Contardo2025, Lovell_2025_inference}, 
constraining cosmology and astrophysical processes with observed galaxy scaling relations \citep{Busillo2023,Busillo2025,Tortora2025},
exploring the effects of systematic subgrid parameter variations on the intergalactic medium \citep[e.g.][]{Parimbelli_2023_IGM_param_var, Contreras_2023_IGM_param_var, Tillman_2023b_lyalpha}, circumgalactic medium \citep[e.g.][]{Moser_2022_CGM_param_var, Medlock_2024_fb_energy,Medlock_2024_frb,Medlock_2025_frb, Lau_2024_CGM}, and matter clustering \citep[e.g.][]{Pandey_2023_powerspec, Delgado_2023_powerspec, Gebhardt_2024a}, as well as emulating baryonic properties including HI maps \citep{hassan_2022_HIFlow, Andrianomena_2022_multifield_emu, friedman_2022_HIGlow, Hassan_2023_HI_emu}, galaxy properties \citep{Lovell_2023_NFs,Lovell_2025_inference}, and effects on large-scale matter clustering \citep{Sharma_2024_emu}.

In this work, we use the CAMELS simulations to investigate the impact of baryonic physics on the distribution of matter across four different galaxy formation models and thousands of subgrid parameter variations. In \cite{Gebhardt_2024a}, we investigated the decoupling of baryons from the dark matter component focusing on the baryonic spread metric and its connection to the suppression of the total matter power spectrum. Here we quantify explicitly the impact of baryonic physics on the matter distribution and the back-reaction effect on dark matter at the level of (1) global halo masses, (2) the distribution of mass within individual haloes, and (3) the large-scale clustering of matter. 

This paper is organized as follows: In Section \ref{Methods} we summarize the CAMELS simulations and data products that we use throughout the paper and describe our algorithm to match haloes across hydrodynamic and \textit{N}-body simulations. In Section \ref{Results}, we explore the impact of baryons on the total matter and dark matter components of halo masses, their radial density profiles, and matter power spectra for the fiducial model of each galaxy formation code as well as considering wide cosmological and feedback parameter variations. In Section \ref{Discussion}, we discuss our results in the context of past and future work. In Section \ref{Conclusions} we conclude by summarizing the main findings of this work.

\section{Methods}
\label{Methods}
\subsection{CAMELS Simulations}

\newcommand{\ml}[1]{\begin{tabular}[t]{@{}l@{}}#1\end{tabular}}

\begin{table*}
\centering
\caption{The physical meaning of the four astrophysical parameters ($A_{\rm SN1}$, $A_{\rm SN2}$, $A_{\rm AGN1}$, $A_{\rm AGN2}$) in the SIMBA, IllustrisTNG, ASTRID and Swift-EAGLE suites. The fiducial parameter value in each simulation is normalized to $A_{\rm SN1}$ = $A_{\rm SN2}$ = $A_{\rm AGN1}$ = $A_{\rm AGN2}$ = 1. }
\label{tab:feedback_params}
\renewcommand{\arraystretch}{1.25}
\begin{tabular*}{\textwidth}{@{\extracolsep{\fill}} l l l l l}
\hline
\ml{Simulation} & \ml{$A_{\rm SN1}$} & \ml{$A_{\rm SN2}$} & \ml{$A_{\rm AGN1}$} & \ml{$A_{\rm AGN2}$} \\
\hline
\ml{SIMBA}
  & \ml{Galactic winds:\\ mass loading}
  & \ml{Galactic winds:\\ wind speed}
  & \ml{QSO \& jet-mode BH feedback:\\ momentum flux}
  & \ml{Jet-mode BH feedback:\\ jet speed} \\
\ml{IllustrisTNG}
  & \ml{Galactic winds:\\ energy per unit SFR}
  & \ml{Galactic winds:\\ wind speed}
  & \ml{Kinetic mode BH feedback:\\ energy per unit BH accretion}
  & \ml{Kinetic mode BH feedback:\\ ejection speed / burstiness} \\
\ml{ASTRID}
  & \ml{Galactic winds:\\ energy per unit SFR}
  & \ml{Galactic winds:\\ wind speed}
  & \ml{Kinetic mode BH feedback:\\ energy per unit BH accretion}
  & \ml{Thermal mode BH feedback:\\ energy per unit BH accretion} \\
\ml{Swift\textendash EAGLE}
  & \ml{Galactic winds:\\ energy per unit SFR}
  & \ml{Galactic winds:\\ metallicity dependence}
  & \ml{BH feedback:\\ scaling of Bondi accretion rate}
  & \ml{BH feedback: temperature jump of \\ gas particles in feedback events} \\
\hline
\end{tabular*}
\label{tab:table1}
\end{table*}

In this work, we use simulations from the CAMELS public dataset\footnote{\url{https://camels.readthedocs.io/en/latest/}} \citep{Villaescusa_Navarro_2021, CAMELS_data_release, Ni_2023_CAMELSastrid}. Each simulation has a comoving volume of $(\text{25\,Mpc}\,h^{-1})^{3}$ with $256^{3}$ dark matter particles of mass $6.49 \times 10^{7} (\Omega_{\rm{m}} - \Omega_{\rm{b}})/0.251\,h^{-1}\rm{M_{\odot}}$, and $256^{3}$ gas particles of mass $1.27 \times 10^{7} \Omega_{\rm{b}}/0.049\,h^{-1}\rm{M_{\odot}}$. Throughout this paper, we adopt fiducial cosmological parameters $\Omega_{\rm{m}} = 0.3$, $\Omega_{\rm{b}} = 0.049$, $\sigma_{8} = 0.8$, $h = 0.6711$, but $\Omega_{\rm{m}}$ and $\sigma_{8}$ can vary over a wide range. To quantify back-reaction effects, we compare CAMELS $N$-body simulations to their corresponding hydrodynamic counterparts, where each simulation pair shares the same initial conditions. In this work, we use the galaxy formation models of SIMBA \citep{simba_Dave_2019}, IllustrisTNG \citep{weinberger_2017_TNGpaper, Pillepich_2018_TNGpaper}, ASTRID \citep{bird_2022_astridgalaxys, ni_2022_astridSMBH}, and Swift-EAGLE (\citealt{Schaye_2015_EAGLE, Borrow2023sweagle}, Lovell et al. in prep.), which are described below.

\textsc{SIMBA} \citep{simba_Dave_2019}, the successor to MUFASA \citep{mufasa_Dave_2016}, uses GIZMO \citep{gizmo_Hopkins_2015} to solve hydrodynamics and gravitation dynamics, and implements radiative cooling and photoionization from Grackle-3.1 \citep{grackle_Smith_2016}. Stellar feedback is modeled with two-phase, kinetic galactic winds similar to MUFASA, implementing wind velocities and mass loading factors based on results from the FIRE simulations (\citealt{Muratov_FIRE_paper, DAA_2017}). SMBH growth is implemented with a gravitational torque accretion model \citep{hopkins_and_quataert_2011, DAA_2017_torqueBHaccretion} for cold gas and the Bondi accretion model \citep{Bondi_1952} for hot gas. Feedback from AGN is implemented kinetically (\citealt{DAA_2017_torqueBHaccretion}) for quasar-mode winds and high-speed, collimated jets (heated to the virial temperature of haloes).

IllustrisTNG, the successor to Illustris \citep{Genel_2014_illustris, Vogelsberger_2014_illustris2}, uses Arepo \citep{springel_2010_arepo} to solve the equations of gravity and magnetohydrodynamics, with radiative cooling following \cite{katz_1996_tngcooling}, \cite{wiersma_2009_tngcooling}, and \cite{Rahmati_2013_Hshielding}. Stellar feedback galactic winds follow a kinetic scheme based on \cite{springel_hernquist_2003_stellarfeedback}, which is fully described in \cite{Pillepich_2018_TNGpaper}. The SMBH model builds upon \cite{gadget_Springle_2005}, \cite{sijacki_2007_AGNmodel}, \cite{Vogelsberger_2013_cosmosim}, and \cite{weinberger_2017_TNGpaper}. AGN feedback is implemented in three modes: thermal (high accretion mode), kinetic (low accretion mode), and radiative.

\textsc{ASTRID} uses MP-Gadget and the Smoothed Particle Hydrodynamics (SPH) method, with radiative cooling and heating modeled from \cite{Vogelsberger_2013_cosmosim}. Stellar feedback is implemented kinetically, where wind particles are generated from newly-formed star particles. The CAMELS version of \textsc{ASTRID} \citep{Ni_2023_CAMELSastrid} extends the SMBH model from the ASTRID production run \citep{bird_2022_astridgalaxys,ni_2022_astridSMBH} to include a two-mode SMBH feedback implementation using thermal (high accretion mode) or kinetic (low accretion mode) energy injections into the surrounding gas within a spherical region with radius twice that of the SPH kernel. 

Swift-EAGLE solves equations of hydrodynamics and gravity using the Swift code \citep{Schaller_2024_swift} with the SPHENIX SPH scheme \citep{Borrow_2022_sphenix} and implements a modified version of the Evolution and Assembly of GaLaxies and their Environments (EAGLE) galaxy formation model \citep{Schaye_2015_EAGLE, Crain_2015_EAGLE}. Radiative cooling and heating follows \cite{Wiersma_2009_rad_cooling}, which is implemented element by element. Star formation follows a modified version of \cite{Schaye_2008_star_formation_theory}, with altered metallicity-dependent density and temperature thresholds. Stellar feedback is implemented stochastically following \cite{DallaVecchia2012}. AGN feedback is implemented following \cite{Booth2009} in a single mode with fixed efficiency, where energy proportional to the gas accretion rate of the black hole is injected thermally and stochastically. Both stellar and AGN feedback are modeled as spherically-symmetric injection of thermal energy in discrete events after accumulating enough feedback energy to heat gas up to a predefined minimum temperature to avoid cooling losses. 
The CAMELS Swift-EAGLE simulations (presented in Lovell et. al, {\it in prep.}) implement the same subgrid physics as \citet{Schaller_2024_swift} but recalibrated at lower resolution to maintain consistency with the other CAMELS simulations.

\subsection{Datasets and parameter variations}
Suites of simulations in CAMELS are subdivided into datasets depending on how parameters or initial conditions are varied. In this work, we use the following three simulation sets:

\begin{itemize}
    \item The CV (``cosmic variance'') set contains 27 simulations in which all parameters are held constant at fiducial values, but each are run with different initial conditions.

    \item The 1P (``one parameter'') set contains simulations in which one parameter is varied while the rest are held constant. The initial conditions are the same in each simulation. The full 1P set varies up to 28 parameters depending on the simulation suite \citep{Ni_2023_CAMELSastrid}, each with four variations of the fiducial value. In this work, we use the ``six-parameter 1P set,'' which varies two cosmological parameters ($\Omega_{\rm m}$ and $\sigma_{8}$) and four astrophysical parameters governing the efficiency of SNe and AGN feedback in their respective galaxy formation models. The physical meaning of the four astrophysical parameters in each model is described in Table \ref{tab:table1}.

    \item The LH (``Latin Hypercube'') set contains 1000 simulations simultaneously varying the six parameters and initial conditions, with parameters selected from a Latin Hypercube. 
\end{itemize}

The results presented throughout this paper focus on quantifying the impact of baryonic physics on the distributions of dark matter and total matter. Besides the data presentation papers for each simulation suite (\citealt{Villaescusa_Navarro_2021, Ni_2023_CAMELSastrid}; Lovell et. al, {\it in prep.}), we refer the reader to previous CAMELS papers for a general description of the simulated galaxy samples, including galaxy scaling relations \citep{Villaescusa-Navarro_2022_OneGalaxy,Busillo2023,Lue2025}, colors and luminosity functions \citep{Lovell_2025_inference}, star formation histories \citep{Iyer2025}, and circum-galactic medium properties \citep{Delgado_2023_powerspec,Lau_2024_CGM,Medlock_2024_fb_energy}.

\subsection{Halo matching across simulations}
In this work, we use {\sc SUBFIND} \citep{Springel_2001_SUBFIND} halo catalogs to identify and compute global properties of haloes, defined as a spherical virial region within which the average density is 200 times the critical density of the Universe. We compare haloes between the hydrodynamic and $N$-body simulation pairs in CAMELS accurately by matching their identity based on the number of common dark matter particles, as follows:


\begin{enumerate}
\item Each dark matter particle is given a halo ID based on the identity of the host halo.

\item For a given halo in the $N$-body simulation, we record the IDs of all dark matter particles within the virial radius, $r_{200}$.

\item We then find the same particles (matched by particle ID) in the corresponding hydrodynamic simulation, and compute the mode of their halo ID.

\item The mode of the selected halo IDs in the hydrodynamic simulation indicates the matching halo to which most of the selected dark matter particles belong.

\end{enumerate} 
We perform this computation for all haloes that contain at least 100 dark matter particles and that additionally match in the opposite direction (starting with the hydrodynamic simulation and matching to the $N$-body simulation). This ``cross-matching'' allows us to select only haloes that matched both ways, filtering out haloes that may have merged in one simulation and not the other. We can also further select haloes by the fraction of matched particles (defined when matching from the $N$-body simulation to the hydrodynamic counterpart). 
In this work, we conservatively consider haloes that shared at least $>$60\% of matched particles. Comparing the numbers of matched haloes and recomputing key quantities such as $N$-body-to-hydrodynamic halo mass ratios for varying particle matched thresholds show that our results are not sensitive to this choice given the cross-matching approach between the $N$-body and hydrodynamic simulations.

The small comoving volume of $(\text{25\,Mpc}\,h^{-1})^{3}$ limits the number of matched haloes available in each CAMELS simulation, particularly in the high mass regime. Combining the full CV set of simulations for each suite allows us to robustly quantify the impact of baryonic physics on global halo properties for $M_{\rm halo} \lesssim 10^{14}\,\rm{M_{\odot}}$ (with $>$50 haloes in the mass range $M_{\rm halo} > 10^{13.5}\,\rm{M_{\odot}}$). However, analyses of single-parameter variations using the 1P simulation sets rely on individual volumes and are thus limited to $M_{\rm halo} \lesssim 10^{13}\,\rm{M_{\odot}}$ (with $\sim$20 halos in the range $M_{\rm halo} = 10^{12.5-13}\,\rm{M_{\odot}}$ but only $\sim$4 haloes at higher masses). 
On the other hand, given the constraints in spatial ($\sim$1\,kpc) and dark matter mass ($\sim$10$^{8}\,\rm{M}_{\odot}$) resolution, we limit the analysis of halo radial profiles to $M_{\rm halo} > 10^{11}\,\rm{M_{\odot}}$.

\section{Results}
\label{Results}
The inclusion of baryons and their associated physics directly affects many measurements relevant for cosmological and galaxy evolution analyses.
We first investigate this effect on haloes, including masses and radial profiles, and later explore the effect on the matter power spectrum. For all analyses here, we first compare changes in the total matter distribution between $N$-body simulations and the corresponding hydrodynamic simulations. Subsequently, we compare changes to the dark matter component alone to investigate back-reaction effects in detail. 
Relative to $N$-body simulations, dark matter particle masses are reduced in the hydrodynamic simulations to account for the added baryonic mass. When exploring back-reaction effects on dark matter specifically, we scale up dark matter particle masses in the hydrodynamic simulations by a factor $\Omega_{\rm{m}}/(\Omega_{\rm m} - \Omega_{\rm b})$ to match the particle masses in the $N$-body simulations \citep[e.g.,][]{van_Daalen_2011}. In this way, comparing scaled dark matter masses and radial profiles in hydrodynamic simulations to their $N$-body counterparts provide an unbiased measure of the intrinsic impact of baryonic physics on the distribution of the dark matter component (normalizing out the trivial effect of depositing mass into baryonic particles).


\subsection{Halo masses and baryon fractions}

\subsubsection{Fiducial simulations}

\begin{figure*}
	\includegraphics[width=\textwidth]{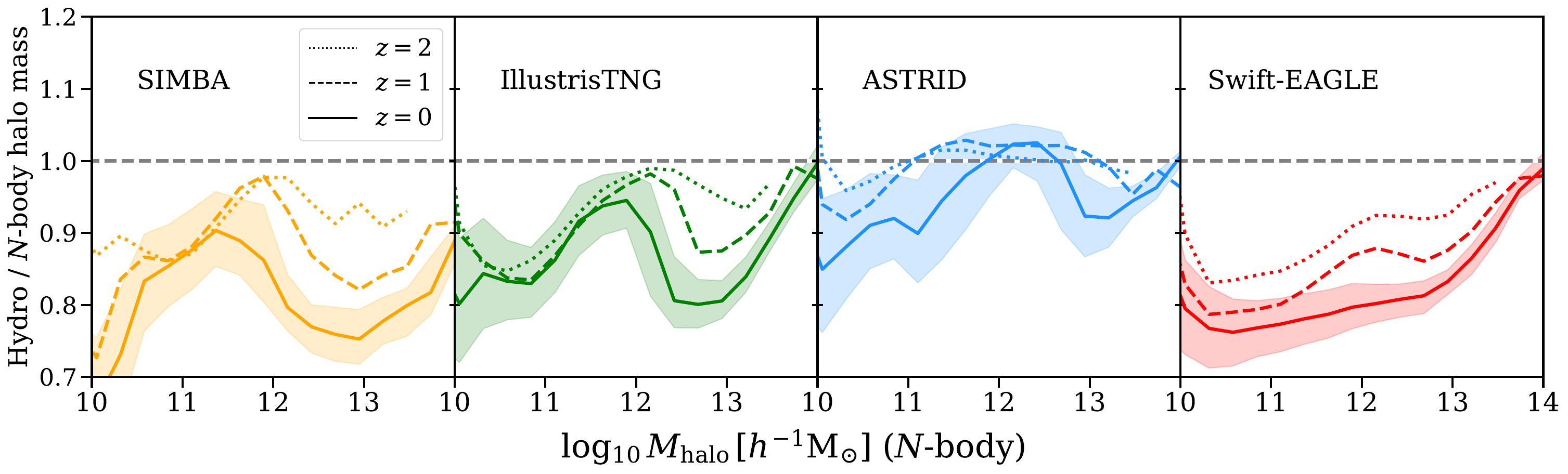}
    \vspace{-0.2in}
    \caption{Ratio of matched hydrodynamic to $N$-body halo total masses at $z=0$ (solid lines), $z=1$ (dashed lines), and $z=2$ (dotted lines) as a function of $N$-body halo mass for all matched haloes in the CV set for SIMBA, IllustrisTNG, ASTRID, and Swift-EAGLE (from left to right). Lines represent the median value in each mass bin, while the shaded region represents the $25^{\rm th}$ to $75^{\rm th}$ percentile range at $z=0$. Haloes are matched across hydrodynamic and $N$-body simulations with the same initial conditions, and are selected and binned by the $N$-body halo mass. SIMBA and Swift-EAGLE show the greatest reduction in total halo mass relative to $N$-body simulations, followed by IllustrisTNG and ASTRID. Baryonic effects on halo mass are generally greater at later times.}
    \label{fig:total_halo_mass}
\end{figure*}

\begin{figure*}
	\includegraphics[width=\textwidth]{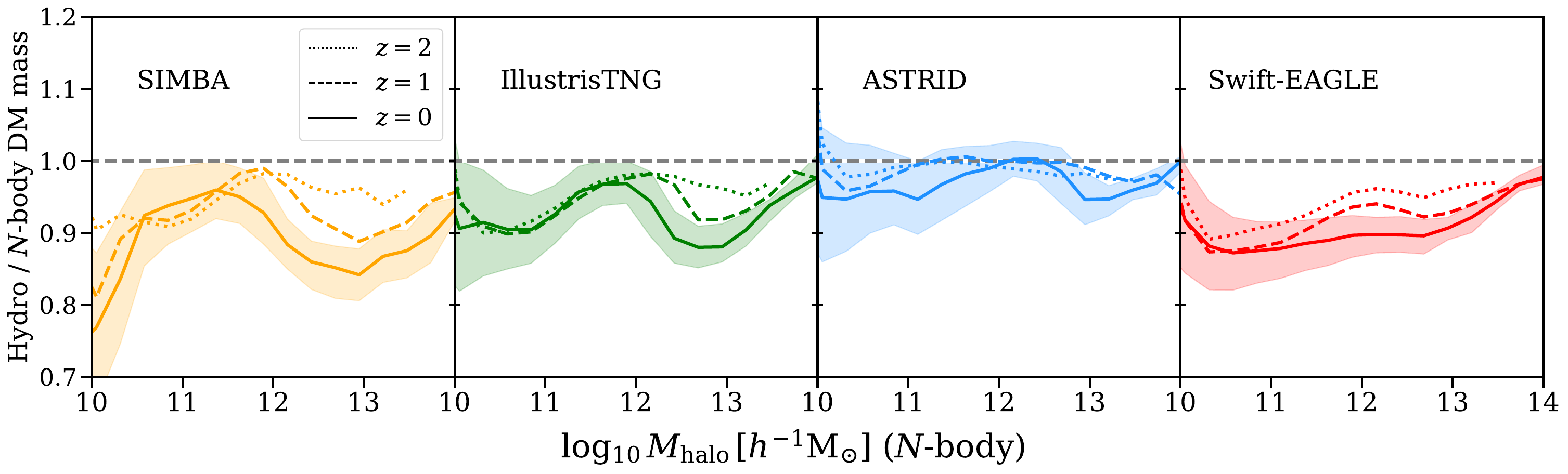}
    \vspace{-0.2in}
    \caption{Same as Figure \ref{fig:total_halo_mass} but for the mass of the dark matter component of haloes. Dark matter particle masses in hydrodynamic simulations are scaled up to be the same as in the $N$-body run, for comparison.
    The dark matter content of haloes in hydrodynamic simulations follows the same general trend as total halo mass relative to $N$-body simulations, but with a lower magnitude of variation. Back-reaction effects generally reduce dark matter halo masses, 
    with stronger impact at later times.
    }
    \label{fig:dm_halo_mass}
\end{figure*}

We begin by looking at how the masses of haloes change with the inclusion of baryons and baryonic physics in the four simulation suites. Figure \ref{fig:total_halo_mass} shows the ratio of matched hydrodynamic to $N$-body halo total masses at $z=0$, $z=1$, and $z=2$ as a function of $N$-body halo mass for all matched haloes in the CV set for SIMBA, IllustrisTNG, ASTRID, and Swift-EAGLE (from left to right). Each line represents the median value in each mass bin, and the shaded region represents the $25^{\rm th}$ to $75^{\rm th}$ percentile range for $z=0$ (the variance is comparable at $z=1$ and $z=2$ and is omitted for clarity). Haloes are selected and binned based on the mass of the matched $N$-body halo. 

At $z=0$, halo masses in all models are decreased relative to $N$-body simulations across the full halo mass range, except for ASTRID at $M_{\rm halo} \sim 10^{12}\,\rm{M_{\odot}}$. In SIMBA, IllustrisTNG, and ASTRID, halo masses are most reduced at $M_{\rm halo} < 10^{11}\,\rm{M_{\odot}}$ and $M_{\rm halo} \sim 10^{13}\,\rm{M_{\odot}}$, while haloes with $M_{\rm halo} \sim 10^{12}\,\rm{M_{\odot}}$ and $10^{14}\,\rm{M_{\odot}}$ are comparably less affected. In Swift-EAGLE, the mass reduction is more uniform across the full mass range, but is generally less significant at higher masses. In general, halo masses are progressively reduced from $z=2 \rightarrow 0$ in all models. In SIMBA and IllustrisTNG, the largest changes occur at $M_{\rm halo} \sim 10^{13}\,\rm{M_{\odot}}$, while the effects are roughly equal across all masses in Swift-EAGLE.

The back-reaction effect of baryons on dark matter in haloes is shown in Figure \ref{fig:dm_halo_mass}, which is similar to Figure \ref{fig:total_halo_mass} but now for the mass of the dark matter component in haloes. For haloes in hydrodynamic simulations, dark matter particle masses are scaled up to be the same as in the corresponding $N$-body simulation, such that the plotted ratio would equal 1.0 at all masses in the absence of back-reaction effects on the dark matter component. In general, the effects on dark matter follow the same trends and halo mass dependence as for the total matter content at $z=0$: dark matter masses are generally reduced relative to their $N$-body counterparts, with the strongest effects in SIMBA and Swift-EAGLE, and the weakest effects in ASTRID. The back-reaction effect (reduction of dark matter mass) increases over time (from $z=2 \rightarrow 0$) in all models.

\begin{figure*}
	\includegraphics[width=\textwidth]{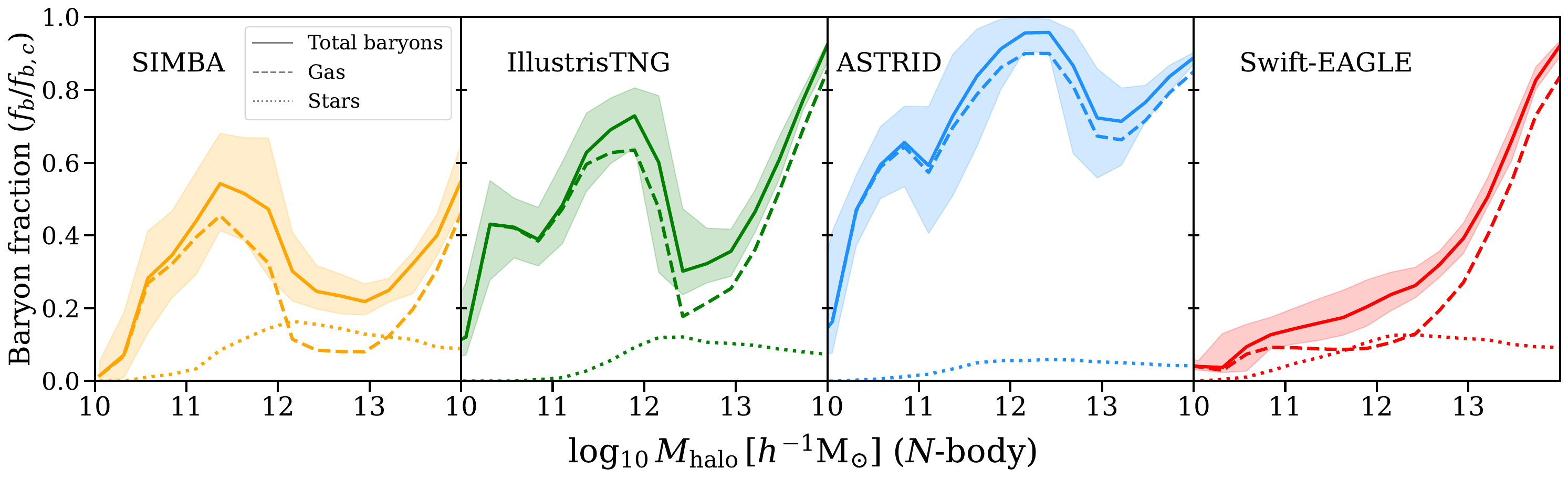}
    \vspace{-0.2in}
    \caption{Halo baryon fraction relative to the cosmic mean at $z=0$ as a function of $N$-body halo mass for all matched haloes in the CV sets of SIMBA, IllustrisTNG, ASTRID, and Swift-EAGLE (from left to right). The solid line represents the median value in each mass bin, while the shaded region represents the $25^{\rm th}$ to $75^{\rm th}$ percentile range. The dashed and dotted lines represent the gas and stellar components, respectively. Overall, SIMBA, IllustrisTNG, and ASTRID show qualitatively similar trends (albeit with different magnitudes) as a function of halo mass while Swift-EAGLE shows significantly lower baryon fractions in lower mass haloes that increase in proportion to halo mass.}

    \label{fig:fb}
\end{figure*}

To conclude our analysis of global halo properties in the fiducial models, Figure \ref{fig:fb} shows the baryon fraction (relative to the mean cosmic baryon fraction, $f_{b,c} \equiv \Omega_{\rm b}/\Omega_{\rm m}$) of hydrodynamic haloes as a function of matched $N$-body halo mass at $z=0$. In this case, the dashed and dotted lines show the fraction of halo mass in gas and stars. SIMBA, IllustrisTNG, and ASTRID predict progressively higher overall baryon fractions, respectively, but show qualitatively similar trends: low baryon fraction at halo masses where stellar feedback is strong relative to the gravitational potential ($M_{\rm halo} < 10^{11}\,\rm{M_{\odot}}$), a peak in baryon fraction at $M_{\rm halo} \sim 10^{11.5-12}\,\rm{M_{\odot}}$, followed by a minimum at $M_{\rm halo} \sim 10^{13}\,\rm{M_{\odot}}$, where feedback from AGN becomes most efficient, and finally rising baryon fractions at higher masses ($M_{\rm halo} > 10^{13}\,\rm{M_{\odot}}$) where the gravitational potential is strong enough to retain baryons in spite of AGN feedback. 
In contrast, Swift-EAGLE yields significantly different baryon fractions systematically increasing across the halo mass range $M_{\rm halo} = 10^{10}\,\rm{M_{\odot}} \rightarrow 10^{14}\,\rm{M_{\odot}}$, demonstrating the impact of different feedback implementations.
In all models, gas generally dominates the baryonic content of haloes across the halo mass range. Interestingly, the fraction of mass in stars peaks at higher halo mass ($M_{\rm halo} \sim 10^{12-13}\,\rm{M_{\odot}}$) than the baryon fraction, coinciding with the local minimum in halo gas fraction in SIMBA and IllustrisTNG (the stellar fraction becomes dominant in SIMBA at $M_{\rm halo} \sim 10^{12-13}\,\rm{M_{\odot}}$ owing to the efficient ejection of gas by AGN feedback). 
The baryon fraction in general correlates with total and dark matter mass in hydrodynamic haloes relative to their $N$-body counterparts across all models, with higher baryon fractions indicating a greater retention of mass.

\subsubsection{Impact of parameter variations on halo masses}

\begin{figure*}
	\includegraphics[width=\textwidth]{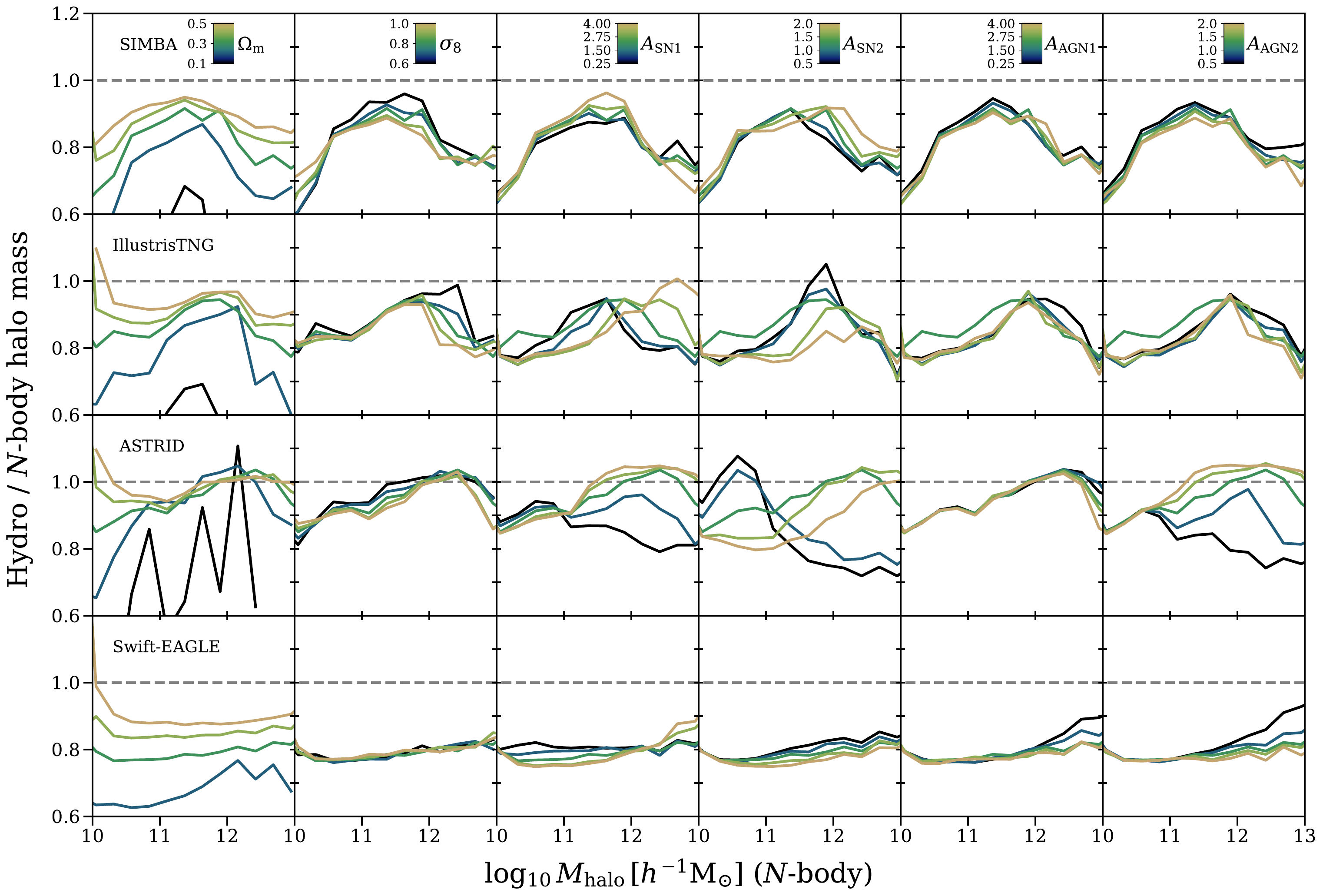}
    \vspace{-0.2in}
    \caption{Ratio of hydrodynamic to $N$-body halo total masses at $z=0$ as a function of $N$-body halo mass for all matched haloes in the six-parameter 1P set for SIMBA, IllustrisTNG, ASTRID, and Swift-EAGLE (same as Figure~\ref{fig:total_halo_mass} but for cosmological and feedback parameter variations). Each row corresponds to a different simulation suite, and each column corresponds to variations of a different parameter, as indicated in the first row and column. Lines of different colors in each panel show the median halo mass ratio for simulations varying only the corresponding parameter, as indicated by the color bar. 
    Increasing $\Omega_{\rm m}$ (at fixed $\Omega_{\rm b}$) generally increases the mass of hydrodynamic haloes relative to $N$-body simulations (suggesting weaker feedback efficiency with lower baryon to dark matter content), while increasing $\sigma_{8}$ has contrasting effects depending on feedback model. Increasing AGN feedback efficiency generally decreases halo mass at $M_{\rm halo} > 10^{12}\rm{M_{\odot}}$ (but see $A_{\rm{AGN2}}$ in ASTRID). Increasing SNe feedback efficiency generally decreases halo mass at $M_{\rm halo} \lesssim 10^{11} \rm{M_{\odot}}$ but can have the opposite effect in higher mass haloes due to non-linear interaction of stellar and AGN feedback, with significant variation between models.}
    \label{fig:halo_mass_variations}
\end{figure*}

As shown above, the masses of haloes in hydrodynamic simulations are affected by the presence and related physics of baryonic matter. Next, we explore the impact of varying cosmological and feedback parameters on the global evolution of haloes using the 1P set in each simulation suite. Figure \ref{fig:halo_mass_variations} is similar to Figure \ref{fig:total_halo_mass} but now examines the effects of parameter variations on halo masses. In each panel, color represents the parameter value (as shown in the color bar), with each column corresponding to a different parameter and each row corresponding to a different simulation suite.
We note that our analyses of single-parameter variations are limited to $M_{\rm halo} < 10^{13}\,\rm{M_{\odot}}$ given the limited number of massive haloes available in individual 1P simulations.

Increasing $\Omega_{\rm m}$ (at fixed $\Omega_{\rm b}$) benefits the growth of haloes in the hydrodynamic simulations more than their $N$-body counterparts (possibly indicating weaker feedback efficiency with lower baryon to dark matter content), as the mass ratio increases at all halo masses in each model. Increasing $\sigma_{8}$ generally decreases halo masses, but the specific impact depends on feedback model: In SIMBA and IllustrisTNG, higher $\sigma_{8}$ values correlate with lower hydrodynamic masses relative to $N$-body at higher masses, while in ASTRID, higher values of $\sigma_{8}$ correlate with a decrease of hydrodynamic halo mass relative to $N$-body primarily at lower and intermediate masses ($M_{\rm halo} \sim 10^{11.5}\,\rm{M_{\odot}}$). In Swift-EAGLE, there is generally no impact of varying $\sigma_{8}$.

The $A_{\rm{SN1}}$ parameter generally represents the energy per unit star formation rate in galactic winds in all models. In general, increasing $A_{\rm{SN1}}$ tends to suppress the growth of low-mass haloes ($M_{\rm halo} < 10^{11-11.5}\,\rm{M_{\odot}}$) in hydrodynamic simulations relative to $N$-body simulations while enabling more efficient growth at higher halo masses (but see opposite trend in SIMBA). Results from \cite{Ni_2023_CAMELSastrid} show that greater values of $A_{\rm{SN1}}$ corresponded to a reduced black hole mass function in IllustrisTNG and ASTRID. This suggests that the increased stellar feedback efficiency inhibits the growth of SMBHs to allow for more efficient growth of higher mass haloes that would otherwise act to prevent growth via AGN feedback. 

In SIMBA, IllustrisTNG, and ASTRID, the $A_{\rm{SN2}}$ parameter controls the speed of galactic winds. Increasing $A_{\rm{SN2}}$ generally reduces the masses of lower-mass hydrodynamic haloes, and increases mass for higher-mass haloes (qualitatively similar effects to $A_{\rm{SN1}}$ in some cases, but with significant scatter). Increasing $A_{\rm{SN2}}$ in Swift-EAGLE (metallicity dependence of stellar feedback) slightly decreases hydrodynamic halo growth relative to their $N$-body counterparts across the full halo mass range.

More efficient AGN feedback in SIMBA can reduce halo masses relative to $N$-body simulations across the full halo mass range (where $A_{\rm{AGN1}}$ is the momentum flux and $A_{\rm{AGN2}}$ is jet speed).
Increasing AGN feedback efficiency generally decreases halo mass at high masses ($M_{\rm halo} > 10^{12}\rm{M_{\odot}}$) in Swift-EAGLE, and to a lesser extent, IllustrisTNG. In ASTRID, increasing $A_{\rm{AGN2}}$ (energy per unit black hole accretion in the thermal mode) strongly promotes the growth of haloes relative to $N$-body, where haloes with $M_{\rm halo} > 10^{11}\rm{M_{\odot}}$ tend to be more massive than their $N$-body counterparts when $A_{\rm{AGN2}}$ = 2.0. \cite{Ni_2023_CAMELSastrid} found that increasing $A_{\rm{AGN2}}$ strongly reduced the population of black holes, resulting in a decrease in the total energy injected by AGN feedback, corresponding to an increase in galaxy stellar mass in qualitative agreement with results presented here.



\subsection{Halo profiles}

We explore the impact of baryonic physics on the structure of haloes in the form of radial profiles of mass density and enclosed mass as a function of distance to the center of the halo. We calculate the densities and integrated mass in 30 logarithmically spaced radial bins from $0.01 < r/r_{200} < 10$ for all matched haloes in all hydrodynamic and $N$-body simulations. For haloes in the hydrodynamic simulations, radial bin sizes are calculated based on the matched $N$-body halo radius ($r_{200}$ typically differs across matched haloes). In this way, density profiles and enclosed masses are computed for identical radial bins in physical units, enabling a direct one to one comparison between matched haloes in hydrodynamic and $N$-body simulations. Additionally, when computing dark matter profiles in the hydrodynamic simulations, we scale up dark matter particles masses to match the particle masses in the $N$-body simulations for direct comparison.

\subsubsection{Halo profiles in fiducial simulations}

\begin{figure*}
	\includegraphics[width=\textwidth]{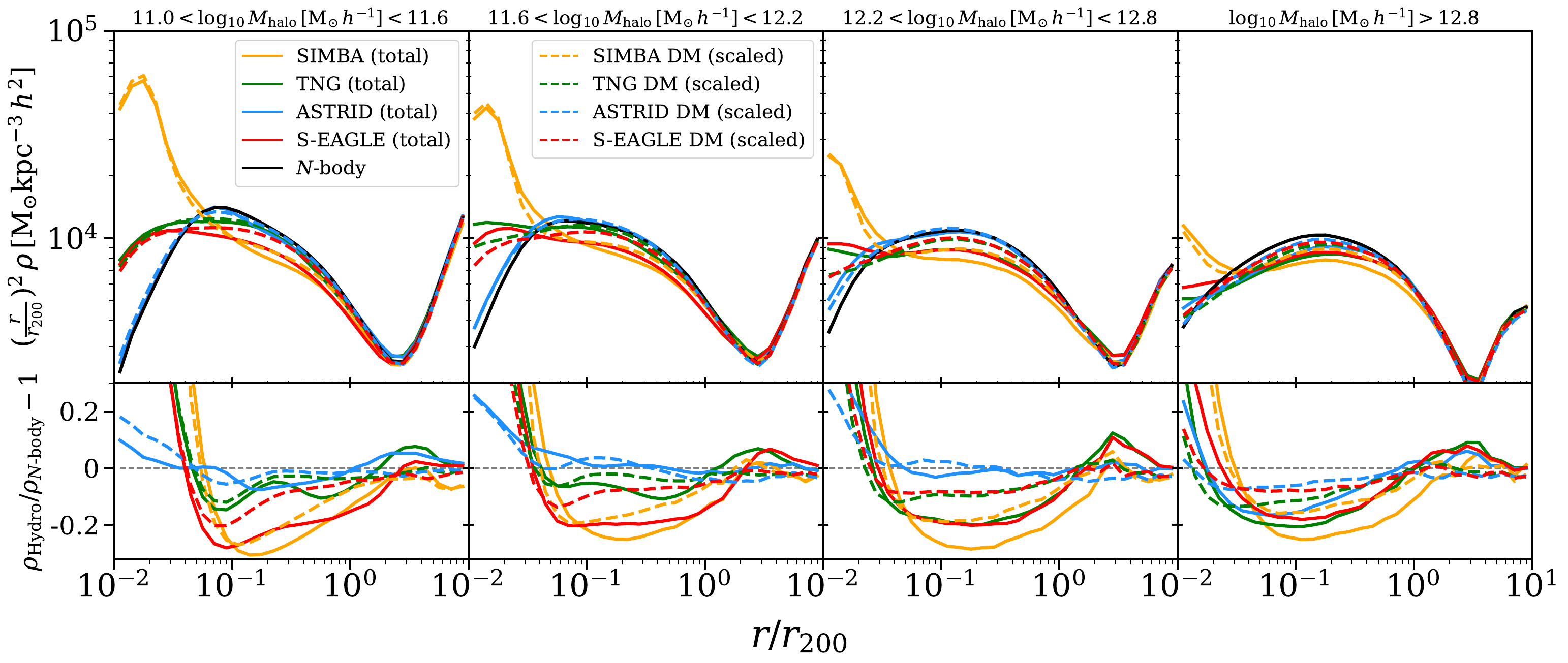}
    \vspace{-0.2in}
    \caption{Total matter (solid lines) and dark matter (dashed lines) density profiles for matched haloes within different mass ranges (increasing from left to right) at $z=0$ in the SIMBA (orange), IllustrisTNG (green), ASTRID (blue), Swift-EAGLE (red) and $N$-body (black) CV simulations (upper panels) and relative difference between each hydrodynamic profile to the $N$-body profile (lower panels) as a function of radial distance from the center of the halo (relative to the virial radius $r_{200}$ of the matched halo in the $N$-body simulation). Density profiles are multiplied by the square of the radial distance in each bin in units of $r_{200}$ to visually enhance the difference between profiles. Haloes are selected based on mass in $N$-body simulations and hydrodynamic haloes are the corresponding matched counterparts of these $N$-body haloes. When calculating dark matter density profiles in hydrodynamic simulations, dark matter particle masses are scaled up to be the same as in the $N$-body run such that the relative difference (bottom panels) for dark matter profiles (dashed lines) would be zero at all radii in the absence of back-reaction effects. Generally, both total and dark matter densities in hydrodynamic haloes are greatly increased at halo centers ($r<0.05\,r_{200}$) relative to $N$-body simulations, while densities are decreased between $r \sim 0.1$--$1\,r_{200}$, with significant differences across galaxy formation models. 
    }
    \label{fig:total_den_pro}
\end{figure*}

The upper panels of Figure \ref{fig:total_den_pro} show total matter and dark matter density profiles for matched haloes within different mass ranges (increasing from left to right) at $z=0$ for fiducial simulations (CV sets) in each galaxy formation model. The lower panels show the fractional difference between the hydrodynamic and $N$-body halo densities in each mass range. 

Generally, both total and dark matter density profiles in hydrodynamic haloes follow qualitatively similar trends, greatly increasing at halo centers ($r<0.05\,r_{200}$) relative to $N$-body simulations owing to radiative cooling while decreasing between $r \sim 0.1$--$1\,r_{200}$ owing to feedback. 
Additionally, the total matter density in hydrodynamic haloes can be larger than in $N$-body simulations on scales beyond the ``splashback radius'' ($r \gtrsim 2.5\, r_{200}$; see e.g. \citealt{Diemer_2014_splashback, Adhikari_2014_splashback, More_2015_splashback, Mansfield_2017_splashback}), likely connected to the distribution of gas ejected by stellar and AGN feedback on scales beyond the virial radius  \citep{Ayromlou_2022_closureradius,Gebhardt_2024a} in addition to back-reaction effects. 
Quantitative differences in radial profiles are clearly seen between models, but tend to decrease for the most massive haloes (right panel). In particular, the reduction of density in intermediate regions gets weaker in SIMBA and Swift-EAGLE, while it gets stronger in IllustrisTNG and ASTRID when considering higher-mass haloes. Overall, ASTRID shows the smallest differences between hydrodynamic and $N$-body matter density profiles across the full halo mass range, consistent with weaker feedback less effective at redistributing mass in haloes compared to the other models.

\begin{figure*}
	\includegraphics[width=\textwidth]{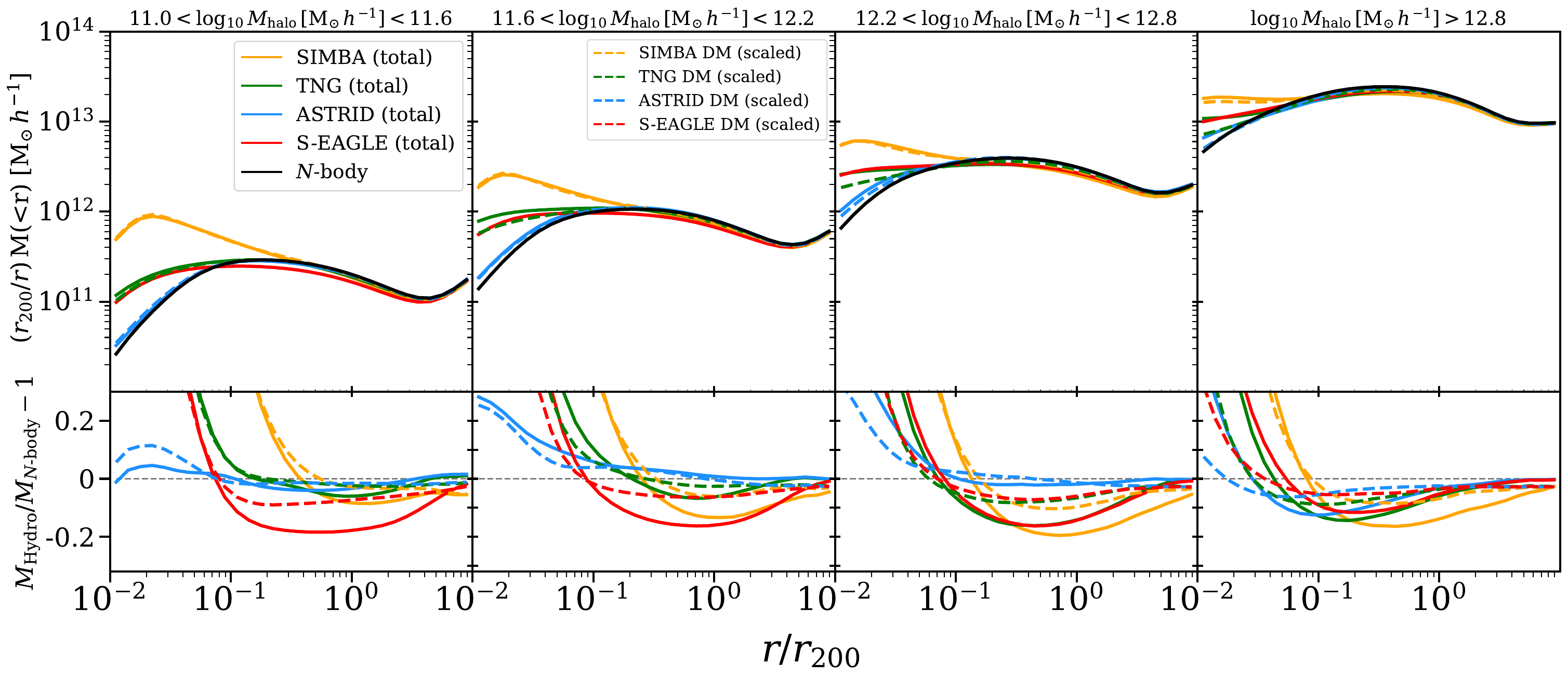}
    \vspace{-0.2in}
    \caption{Same as Figure \ref{fig:total_den_pro} for the enclosed total matter (solid lines) and dark matter (dashed lines) as a function of radial distance for matched haloes within different mass ranges (increasing from left to right) at $z=0$ in the SIMBA (orange), IllustrisTNG (green), ASTRID (blue), Swift-EAGLE (red) and $N$-body (black) CV simulations (upper panels) and relative difference between each hydrodynamic profile to the $N$-body profile (lower panels) as a function of matched $N$-body radial distance. Enclosed mass profiles are divided by the radial distance in units of $r_{200}$ to visually enhance the difference between profiles. Haloes are selected based on mass in $N$-body simulations and hydrodynamic haloes are the corresponding matched counterparts of these $N$-body haloes. As in Figure \ref{fig:total_den_pro}, dark matter particle masses in hydrodynamic simulations are scaled up to be the same as in the $N$-body runs such that the relative difference (bottom panels) for dark matter profiles (dashed lines) directly quantify the back-reaction effect of baryons on dark matter. Enclosed mass profiles show qualitatively similar results as density profiles: the inclusion of baryons results in more mass concentrated in the centers of haloes, while the enclosed mass is reduced farther out relative to $N$-body counterpart haloes.}
    \label{fig:total_mass_pro}
\end{figure*}

We show enclosed mass profiles for total matter and (scaled) dark matter in Figure \ref{fig:total_mass_pro}, which is similar to Figure \ref{fig:total_den_pro} but now for the enclosed mass as a function of radius. Enclosed mass profiles differ the most between models at small radii ($r\lesssim 0.1 \,r_{200}$) and lower mass haloes, with the differences being smaller in the most massive haloes. As expected from the density profiles, SIMBA haloes contain the most mass at inner radii across the halo mass range while having the smallest total enclosed mass at large distances ($r\gtrsim r_{200}$) relative to $N$-body simulations. In contrast, ASTRID enclosed mass profiles are the most similar to $N$-body simulations. At higher masses, we see a better agreement between the models. In general, the (scaled-up) integrated dark matter mass is lower than the total matter mass at small radii, exceeds it at intermediate distances ($r \sim 0.1$–$1,r_{200}$), and converges with it at several times the virial radius. The dark matter profiles in hydrodynamic simulations are significantly different from the dark matter profiles in the $N$-body simulations, indicating very strong back-reaction effects.

\subsubsection{Impact of parameter variations on halo profiles}

\begin{figure*}
	\includegraphics[width=\textwidth]{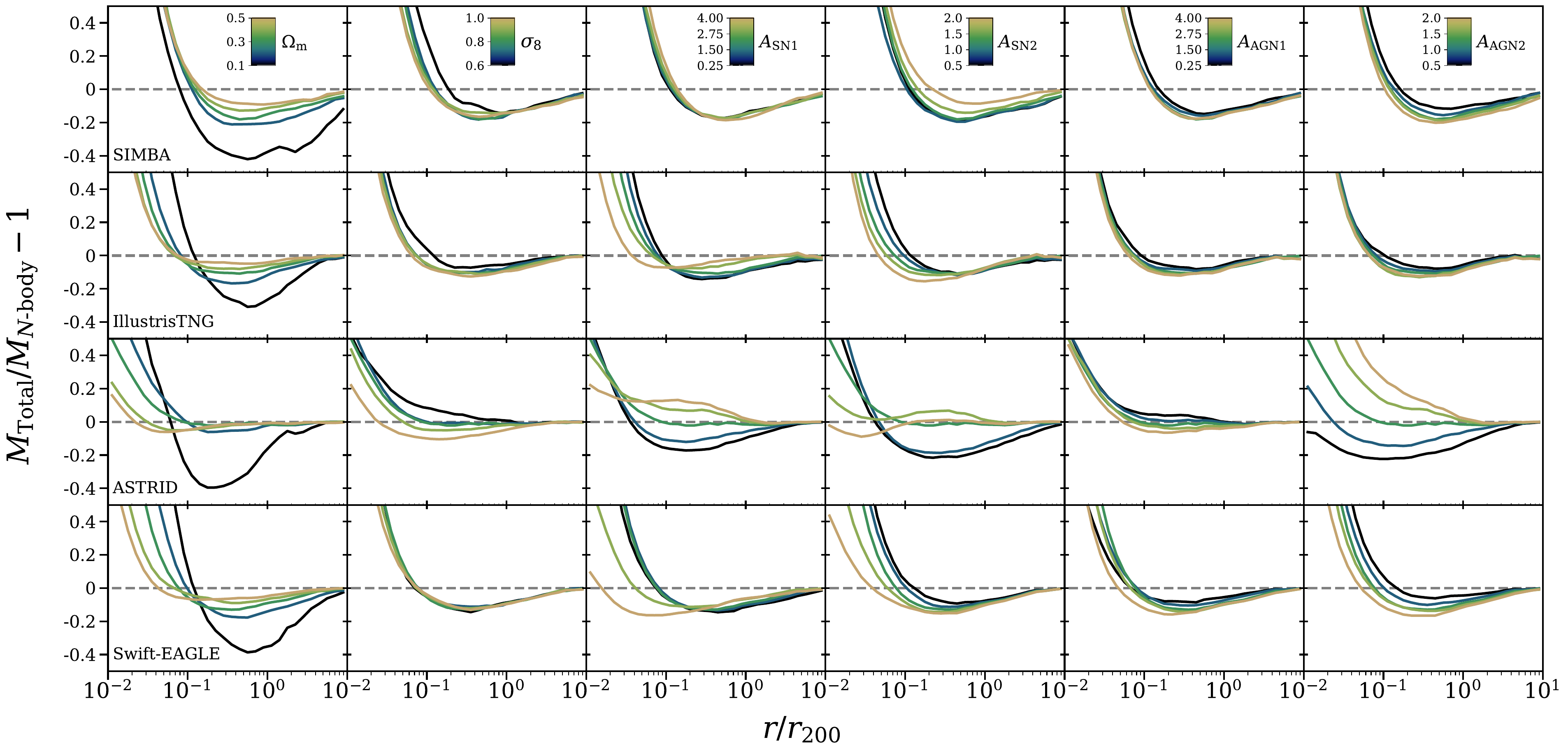}
    \vspace{-0.2in}
    \caption{Relative difference between hydrodynamic and $N$-body total enclosed mass profiles at $z=0$ for all matched haloes with $M_{\rm halo} > 10^{12}\,\rm{M_{\odot}}$ in the six-parameter 1P set for SIMBA, IllustrisTNG, ASTRID, and Swift-EAGLE (similar to the solid lines in the bottom panels of Figure \ref{fig:total_mass_pro} but now for cosmological and feedback parameter variations). Each row corresponds to a different simulation suite, and each column corresponds to variations of a different parameter, as noted in the first row and column. Lines of different colors in each panel indicate the relative difference between enclosed mass profiles for simulations varying only the corresponding parameter, as indicated by the color bar. Halo enclosed mass profiles depend strongly on cosmology at fixed baryonic physics, and the impact of feedback on the distribution of matter in haloes is very sensitive to galaxy formation model and subgrid parameters.}
    \label{fig:total_mass_pro_variations}
\end{figure*}

Next, we investigate the impact of parameter variations on halo mass profiles using the 1P set in each simulation suite. Figure \ref{fig:total_mass_pro_variations} shows the relative difference between hydrodynamic and $N$-body total cumulative mass profiles at $z=0$ for all matched haloes with $M_{\rm halo} > 10^{12}\,\rm{M_{\odot}}$, similar to the solid lines in the bottom panels of Figure \ref{fig:total_mass_pro} but now for cosmological and feedback parameter variations. Varying $\Omega_{\rm m}$ at fixed baryonic physics has an intrinsic effect on the enclosed mass profiles of haloes. While the enclosed mass generally increases at $r \lesssim 0.1\, r_{200}$ and decreases at $r\gtrsim 0.1\, r_{200}$ in hydrodynamic haloes relative to $N$-body simulations, increasing $\Omega_{\rm m}$ (at fixed $\Omega_{\rm b}$) tends to reduce these effects. In other words, hydrodynamic simulations with lower baryonic to dark matter content (lower $\Omega_{\rm b}$/$\Omega_{\rm m}$) are more similar to $N$-body simulations, as expected. At farther radii ($r \sim 0.2$--$10\,r_{200}$), increasing $\Omega_{\rm m}$ increases the enclosed masses of hydrodynamic haloes in SIMBA, IllustrisTNG and Swift-EAGLE relative to $N$-body. On the other hand, increasing $\sigma_{8}$ 
tends to reduce the enclosed masses of all hydrodynamic haloes relative to $N$-body simulations. 
Increasing the SNe feedback parameters ($A_{\rm SN1}$ and $A_{\rm SN2}$) shows qualitatively similar results as increasing $\Omega_{\rm m}$ in most simulations, likely due to reduced star formation (see \citealt{Ni_2023_CAMELSastrid}).

Varying AGN parameters is not as significant, except for changes in $A_{\rm AGN2}$ in ASTRID (energy per unit accretion in thermal mode) and Swift-EAGLE (temperature jump of gas particles in feedback events). Increasing $A_{\rm AGN2}$ results in higher enclosed mass at most radial distances relative to $N$-body in ASTRID, indicating strong self-regulation of black holes in the thermal feedback model reducing the overall ejection of gas from haloes. In contrast, Swift-EAGLE shows the opposite effect with $A_{\rm AGN2}$, ejecting gas from haloes more efficiently as the magnitude of the temperature jump due to AGN feedback increases. Appendix \ref{appendix1} shows similar results for the effects of parameter variations on total and dark matter profiles. 

\begin{figure*}
	\includegraphics[width=\textwidth]{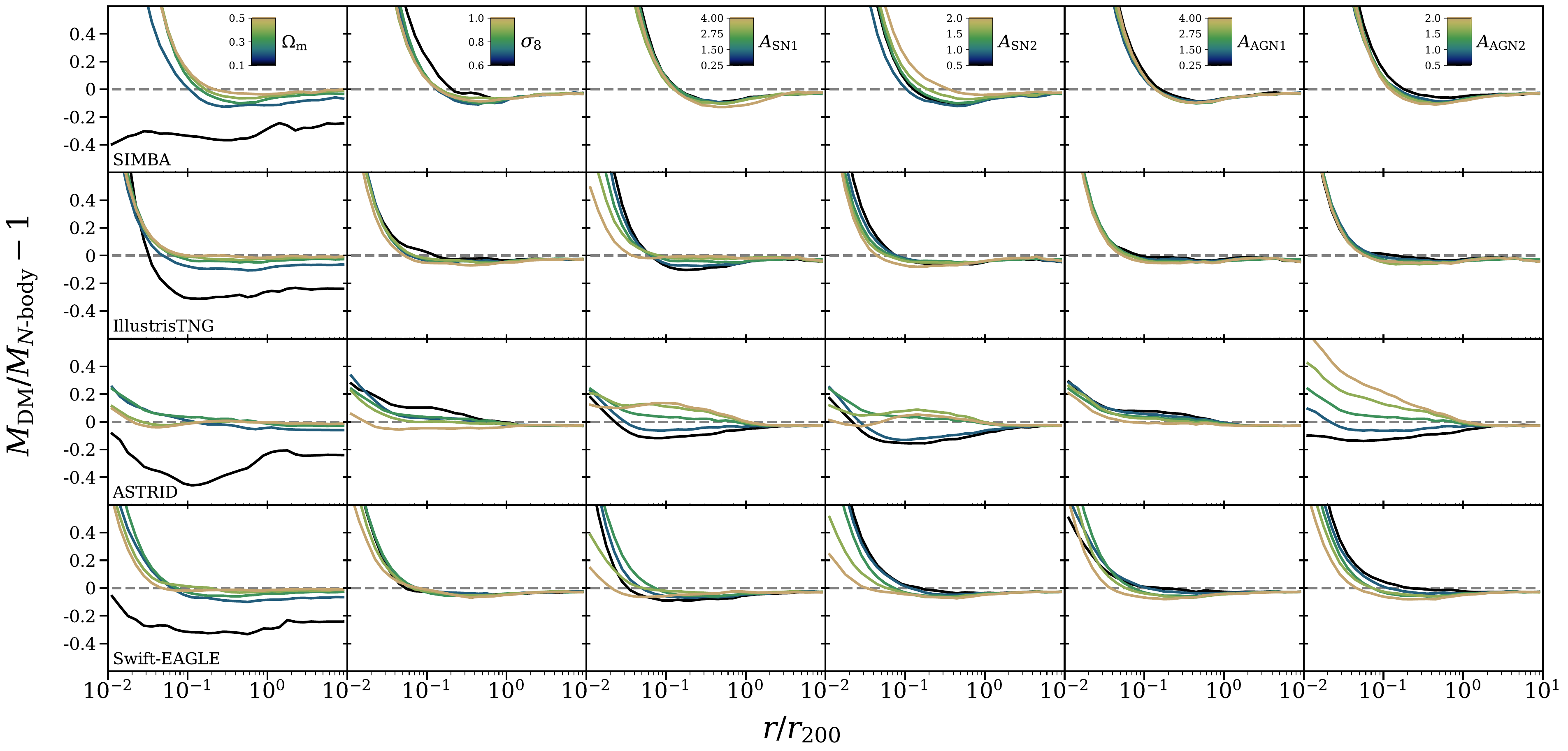}
    \vspace{-0.2in}
    \caption{Same as Figure \ref{fig:total_mass_pro_variations} for the relative difference between hydrodynamic and $N$-body dark matter enclosed mass profiles at $z=0$ for all matched haloes with $M_{\rm halo} > 10^{12}\,\rm{M_{\odot}}$ in the six-parameter 1P set for SIMBA, IllustrisTNG, ASTRID, and Swift-EAGLE. Dark matter particle masses in hydrodynamic simulations are scaled up to be the same as in the $N$-body runs such that the relative difference in enclosed dark matter mass profiles would be zero at all radii in the absence of back-reaction effects. Each row corresponds to a different simulation suite, and each column corresponds to variations of a different parameter, as indicated by the color bar. Variations in dark matter mass profiles generally follow the model-dependent impact of cosmological and feedback parameter variations for total matter density profiles (Figure \ref{fig:total_mass_pro_variations}), with substantial back-reaction effects.}
    \label{fig:dm_mass_pro_variations}
\end{figure*}

Figure \ref{fig:dm_mass_pro_variations} shows the back-reaction effect on halo dark matter profiles as a function of cosmological and astrophysical parameters, similar to Figure \ref{fig:total_mass_pro_variations} but only for the dark matter component. As before, the masses of dark matter particles in the hydrodynamic simulations are scaled up to match those of the $N$-body simulation, for comparison. Most trends with parameter variations are similar to those seen for the total enclosed matter profiles, but with lower amplitude of variation for the dark matter component and more scatter. Additionally, dark matter mass profiles are generally shifted towards the inner regions of the halo, with enclosed dark matter masses in hydrodynamic simulations generally exceeding that of $N$-body simulations at $r \lesssim 0.05\, r_{200}$. 
Simulations with very low total matter content ($\Omega_{\rm m} = 0.1$) represent an interesting exception, with hydrodynamic simulations predicting significantly lower dark matter mass in the inner region of haloes compared to $N$-body in all models except IllustrisTNG. In these extreme cases, feedback dominates over dissipative processes driving an expansion of the dark matter component on all scales.
Clearly, the dark matter mass profiles are strongly affected by both the galaxy formation model and the corresponding parameter variations, indicating non-trivial dark matter back-reaction effects at the level of individual halo profiles.

Overall, both total matter and dark matter profiles show indications of contraction of the inner halo and expansion of the outer halo in hydrodynamic simulations relative to $N$-body for most models and parameter variations. Notable exceptions include the ASTRID model with strong $A_{\rm SN1}$ and $A_{\rm AGN2}$ parameters, increasing the enclosed total and dark matter mass in haloes relative to $N$-body. Dissipative processes may thus overcome feedback depending on the efficiency of stellar and AGN feedback self-regulation.



\subsection{Power spectra}

\subsubsection{Power spectra in fiducial simulations}

\begin{figure*}
	\includegraphics[width=\textwidth]{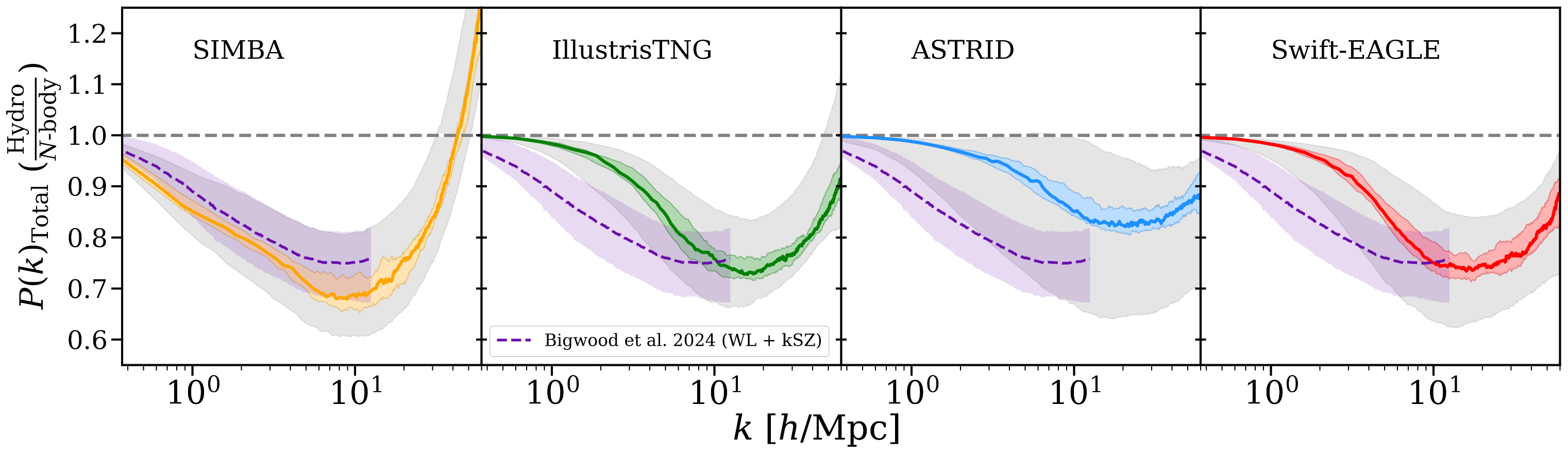}
    \vspace{-0.2in}
    \caption{Ratio of total matter power spectra at $z=0$ from hydrodynamic to $N$-body simulations in the CV sets of SIMBA, IllustrisTNG, ASTRID, and Swift-EAGLE (from left to right). The solid line represents the median value in each $k$ bin over all simulations, while the colored shaded region represents the $25^{\rm th}$ to $75^{\rm th}$ percentile range of variation owing to cosmic variance for the fiducial model in each simulation suite. The gray shaded region represents the $25^{\rm th}$ to $75^{\rm th}$ percentile range for the full LH set varying cosmological and astrophysical parameters in each simulation suite. The dash purple line and shaded region correspond to constraints from \protect\cite{bigwood_2024}, who performed a joint analysis of cosmic shear and kSZ effect measurements. SIMBA shows the greatest suppression of power followed by Swift-EAGLE, IllustrisTNG, and ASTRID, with significant impact of parameter variations in each suite.}
    \label{fig:cv_pk_total}
\end{figure*}

Lastly, we investigate the effect of baryonic physics on the matter power spectrum across a range of spatial scales. We first show the effects of baryons on the total matter distribution before investigating specifically the back-reaction effect on dark matter. Figure \ref{fig:cv_pk_total} shows the ratio of total matter power spectra from hydrodynamic to $N$-body CV set simulations in each suite. Solid lines show median values and colored shaded regions represent the $25^{\rm th}$ to $75^{\rm th}$ percentile range of each CV set, indicating the impact of cosmic variance. For comparison, the gray shaded region represents the $25^{\rm th}$ to $75^{\rm th}$ percentile range for the full LH set varying all cosmological and astrophysical parameters simultaneously. Lastly, we additionally show joint constraints from \cite{bigwood_2024} using DES Y3 cosmic shear and Atacama Cosmology Telescope DR5 kinematic Sunyaev–Zel’dovich (kSZ) measurements (inferred from simulations and thus model dependent).

All four suites show significant suppression of total matter power at $k \sim 10\, h\, \rm Mpc^{-1}$, though the effect is strongest in SIMBA ($\sim$$30\%$ suppression) and weakest in ASTRID ($\sim$$15\%$ suppression). SIMBA, with notably strong AGN feedback, shows nearly 15\% suppression in the matter power spectrum at scales as large as $k \sim 1\, h\, \rm Mpc^{-1}$ (see \citealt{Delgado_2023_powerspec, Gebhardt_2024a}). While the fiducial ASTRID model produces the weakest suppression (see \citealt{Ni_2023_CAMELSastrid}), it is the most affected by parameter variations as seen in the LH set shaded region. SIMBA, with its strong suppression of matter clustering, appears to show the greatest agreement with \cite{bigwood_2024} among the models explored here. Parameter variations in IllustrisTNG, ASTRID, and Swift-EAGLE can help reduce discrepancies between simulations and observations (as indicated by the range of variation in the LH sets), but these models still underpredict the suppression of the matter power spectrum on large scales ($k \lesssim 1\, h\, \rm Mpc^{-1}$) within the parameter space explored here.

\begin{figure*}
	\includegraphics[width=\textwidth]{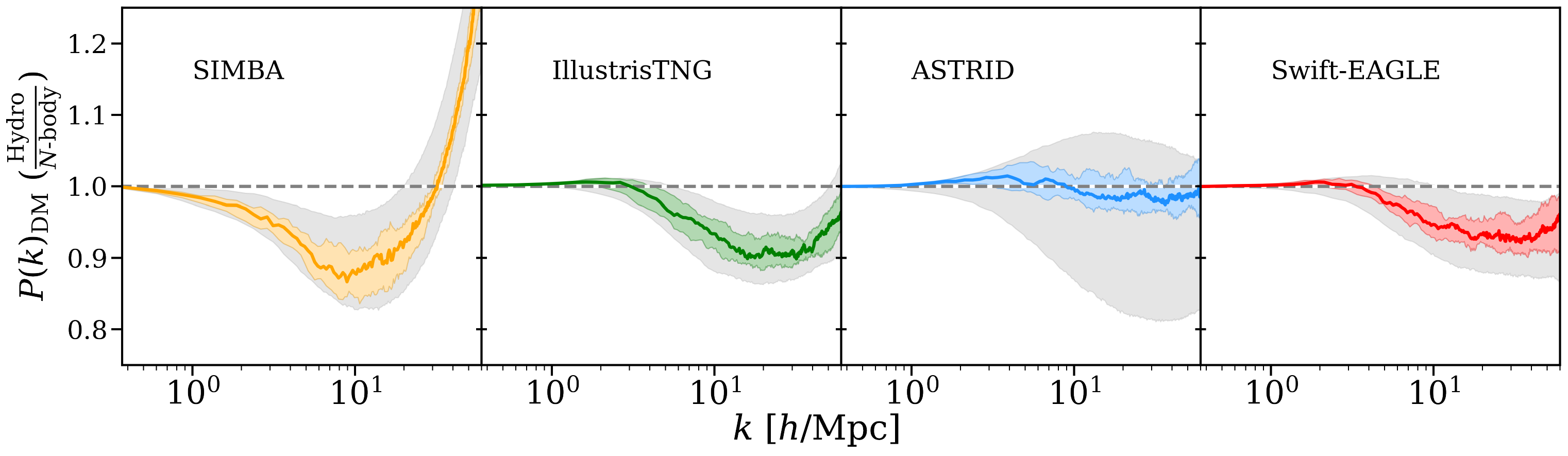}
    \vspace{-0.2in}
    \caption{Same as Figure \ref{fig:cv_pk_total} for the ratio of dark matter power spectra at $z=0$ from hydrodynamic to $N$-body simulations in the CV sets of SIMBA, IllustrisTNG, ASTRID, and Swift-EAGLE (from left to right). The solid line represents the median value in each $k$ bin, while the colored shaded region represents the $25^{\rm th}$ to $75^{\rm th}$ percentile range. The gray shaded region represents the $25^{\rm th}$ to $75^{\rm th}$ percentile for the full LH set. The dark matter power spectrum shows strong back-reaction effects relative to $N$-body simulations, with significant variations across galaxy formation models.}
    \label{fig:cv_pk_br}
\end{figure*}

We show the impact of baryons on dark matter power spectra in Figure \ref{fig:cv_pk_br}, which is similar to Figure \ref{fig:cv_pk_total} but for the dark matter component. As seen in previous results, the baryonic effects on dark matter follow similar trends as total matter. The fiducial models of SIMBA, IllustrisTNG, and Swift-EAGLE all show a strong suppression of dark matter power at $k \sim 10\, h\, \rm Mpc^{-1}$ (up to $\sim$$15\%$ in SIMBA). In contrast, ASTRID shows very weak dark matter back-reaction effects in the fiducial model but shows the greatest change in dark matter power spectra when simulation parameters are varied in the LH set.

\subsubsection{Impact of parameter variations on matter clustering}

\begin{figure*}
	\includegraphics[width=\textwidth]{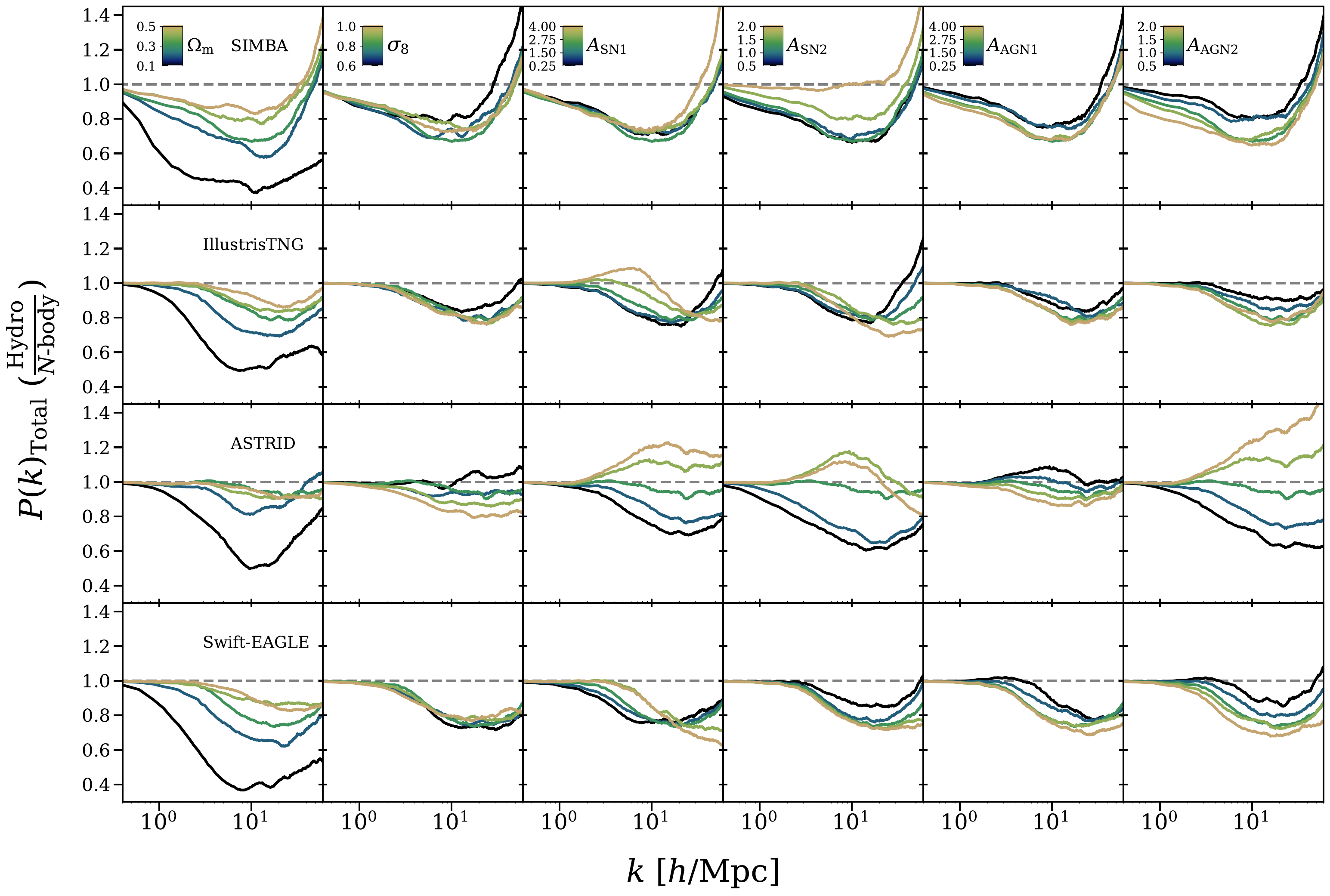}
    \vspace{-0.2in}
    \caption{Ratio of total matter power spectra at $z=0$ from hydrodynamic to $N$-body simulations in the six-parameter 1P sets of SIMBA, IllustrisTNG, ASTRID, and Swift-EAGLE. Each row corresponds to a different simulation suite, and each column corresponds to variations of a different parameter, as noted in the first row and column. Lines of different colors in each panel indicate the total matter power spectrum ratio for simulations varying only the corresponding parameter, as indicated by the color bar. The suppression of power in hydrodynamic simulations increases with lower $\Omega_{\rm m}$ (at fixed $\Omega_{\rm b}$) but models respond differently to changes in $\sigma_8$.  Generally, stronger stellar feedback reduces while stronger AGN feedback enhances the power suppression, but the impact of baryonic physics on matter clustering is very sensitive to galaxy formation model and feedback parameters.}
    \label{fig:pk_variations_tot}
\end{figure*}

We now systematically vary simulation parameters, starting with total matter power spectra in Figure \ref{fig:pk_variations_tot}. Each line corresponds to a hydrodynamic/$N$-body simulation pair with the color denoting the value of the parameter varied in each column (while other parameters are held constant). In all models, higher values of $\Omega_{\rm m}$ (at fixed $\Omega_{\rm b}$) result in less power suppression, due to the reduced impact of feedback in haloes with higher average dark matter to baryonic mass ratios (\citealt{Delgado_2023_powerspec}). Varying $\sigma_{8}$ does not produce systematic changes in the total matter power spectrum for SIMBA and IllustrisTNG, while ASTRID predicts significantly stronger suppression of power at higher $\sigma_{8}$ and Swift-EAGLE predicts the opposite (though weaker) effect.

Most models result in suppression of matter clustering at $k \sim 10\, h\, \rm Mpc^{-1}$ in hydrodynamic simulations relative to $N$-body regardless of feedback parameter variations, but ASTRID can enhance rather than suppress clustering for several 1P simulations.
Increasing the efficiency of SNe feedback generally results in reduced suppression of power, likely due to the non-linear coupling of SNe and AGN feedback that has been observed in previous CAMELS analyses (see e.g. \citealt{Tillman_2023b_lyalpha, Delgado_2023_powerspec, Gebhardt_2024a, Medlock_2024_fb_energy}), but there are significant differences between models. Increasing the galactic wind energy per unit SFR ($A_{\rm{SN1}}$) has a minor effect in SIMBA but drives clear trends in other models. Both IllustrisTNG and Swift-EAGLE increase the suppression of power at $k \gtrsim 20\, h\, \rm Mpc^{-1}$ while reducing the power suppression on larger scales when increasing $A_{\rm{SN1}}$.
In contrast, ASTRID significantly reduces the power suppression at all scales when increasing $A_{\rm{SN1}}$, leading to an actual increase in power at $k$ > $1.5\, h\, \rm Mpc^{-1}$ relative to $N$-body simulations for the highest values of $A_{\rm{SN1}}$.
Increasing the speed of galactic winds ($A_{\rm{SN2}}$ parameter in SIMBA, IllustrisTNG, and ASTRID) generally correlates with reduced power suppression at all scales in SIMBA and ASTRID, while suppression begins to increase at small scales ($k > 20\, h\, \rm Mpc^{-1}$) in IllustrisTNG. 

Increasing the efficiency of AGN feedback generally results in stronger suppression of matter power relative to $N$-body simulations. Increasing either AGN parameter results in greater suppression of power in all simulations with the exception of $A_{\rm{AGN2}}$ in ASTRID (energy per unit black hole accretion of the thermal feedback mode), which has the opposite trend. These results suggest that AGN feedback may be the dominant physical mechanism reducing the clustering of matter across scales, while SNe feedback can inhibit the growth of SMBHs and thus reduce the overall impact (see  \citealt{Singh2022Galactic}).

\begin{figure*}
	\includegraphics[width=\textwidth]{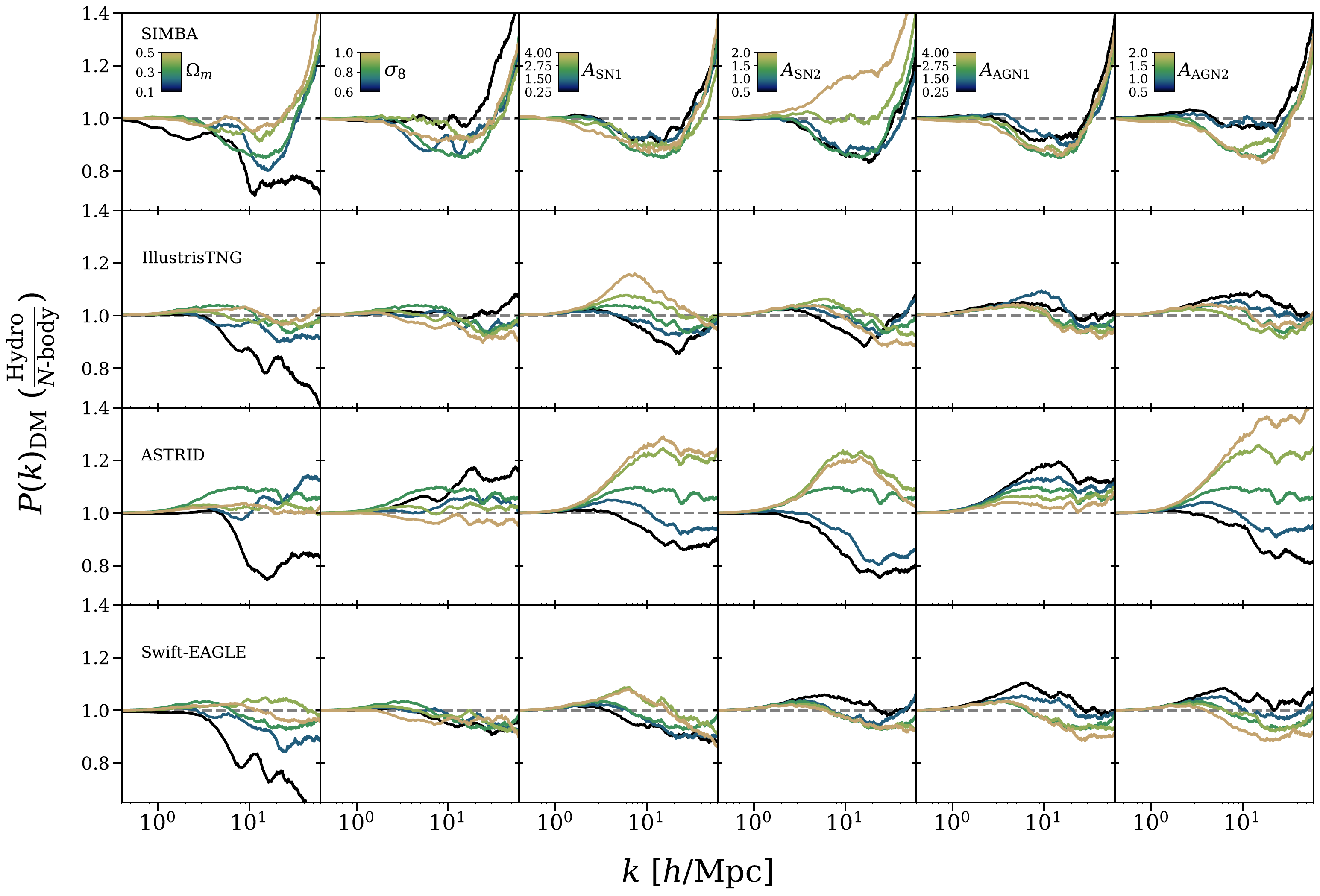}
    \vspace{-0.2in}
    \caption{Ratio of dark matter power spectra at $z=0$ from hydrodynamic to $N$-body simulations in the six-parameter 1P sets of SIMBA, IllustrisTNG, ASTRID, and Swift-EAGLE. Each row corresponds to a different simulation suite, and each column corresponds to variations of a different parameter, as indicated by the colorbar. Variations on cosmological and feedback parameters can very strongly modify the back-reaction of baryons on dark matter power spectra, but the detailed trends depend on the parameter and galaxy formation model.}
    \label{fig:pk_variations}
\end{figure*}

We show the impact of parameter variations on the dark matter power spectrum in Figure \ref{fig:pk_variations}, which is similar to Figure \ref{fig:pk_variations_tot} but now for the ratios of dark matter power spectra in hydrodynamic to $N$-body simulations. Similar to the impact of baryons on individual haloes, the dark matter component follows qualitatively similar trends as the total matter power spectrum, but showing more scatter and less extreme variations relative to $N$-body when changing cosmological and astrophysical parameters. Overall, the dark matter power spectrum depends significantly on the galaxy formation model and specific parameter choices, highlighting the impact of back-reaction effects across a broad range of scales.

\section{Discussion}
\label{Discussion}
It is now well understood that baryons, while comprising a relatively small fraction of the total mass in the Universe, play a key role in shaping the total matter distribution. Unfortunately, the behavior of baryons in simulations is sensitive to the details of the subgrid model. For example, simulations that do not include AGN feedback can produce significantly different results from those that do. \cite{Cui_2012_baryons_on_HMF} compared the masses of haloes from an $N$-body simulation to those from two hydrodynamic simulations: one with only gravitational heating of gas, and one that also included gas cooling, star formation, and stellar feedback. In the former, halo masses were increased in the hydrodynamic simulation on the order of a few percent relative to the corresponding $N$-body simulation depending on the adopted virial radius definition, while in the latter they were increased anywhere between a few percent to 20\% depending on halo mass and radius definition. These simulations did not include AGN feedback, which is now believed to be necessary to solve the overcooling problem in massive galaxies \citep{Fabian_2012_agn_review, Somerville_2015, Crain_2023_SimReview}.

Other works have also investigated the impact of baryonic physics on the growth of dark matter haloes, showing that AGN feedback significantly reduces halo masses. Using Gadget-3 \citep{gadget_Springle_2005} simulations, \cite{Cui_2014} found that AGN feedback reduced halo masses by $\sim 20\%$ at $M_{\rm halo} \sim 10^{12.5}\,\rm{M_{\odot}}$ (using an overdensity of $\Delta_{c} = 500$ to define haloes) relative to $N$-body simulations. Likewise, using simulations from the OWLS project, \cite{Velliscig2014} found a $\sim20\%$ decrease in halo masses at $M_{\rm halo} \sim 10^{13}\,\rm{M_{\odot}}$, with a significant impact up to $M_{\rm halo} \sim 10^{15}\,\rm{M_{\odot}}$ when including AGN feedback. A similar result was reported by \cite{Springel_2018_tng} based on the original IllustrisTNG simulations. More recently, \cite{Sorini_2024_BR} found that halo masses in the flagship IllustrisTNG and MillenniumTNG simulations can be reduced by as much as 20\% depending on the halo mass range considered, and they showed that these results converged across the different simulated boxes with side lengths 50, 100, 300, and 740\,Mpc. Our results are in agreement, and show a very similar trend as a function of halo mass (where halo masses are reduced most where feedback is efficient relative to halo mass, e.g. $M_{\rm halo} \sim 10^{11}\,\rm{M_{\odot}}$ and $10^{13}\,\rm{M_{\odot}}$) in the CAMELS IllustrisTNG run, indicating that the baryonic impact on total halo masses also converges in the smaller $(25\, {\rm Mpc}\, h^{-1})^{3}$ simulations considered here (Figure \ref{fig:total_halo_mass}). Our results are also in agreement with analyses of the original EAGLE simulations \citep{Schaye_2015_EAGLE}, which showed nearly 30\% mass reduction at $M_{\rm halo} \sim 10^{11}\,\rm{M_{\odot}}$ and nearly 20\% reduction at $M_{\rm halo} \sim 10^{12.5}\,\rm{M_{\odot}}$ \citep{schaller_2015_eagle_analysis}.

Crucially, baryons also affect the dark matter content of haloes by modifying the gravitational potential.  We quantified this back-reaction effect at the level of individual haloes across hundreds of model and parameter variations, finding that the dark matter content of haloes follows qualitatively similar trends as total halo masses when comparing hydrodynamic simulations to their $N$-body counterparts. While effects are usually less extreme than for total halo mass (which is affected more directly by the movement of baryons), we find that haloes can end up at $z=0$ with as much as 15-20\% less dark matter (as seen in SIMBA) or halo dark matter masses can also change very little, as is the case in ASTRID. These back-reaction effects represent a huge range of variation across models and assumed feedback efficiencies. 


Our results for the impact of baryonic physics on the matter density profiles of haloes generally agree with previous works. \cite{Sorini_2021_baryons} compared halo density profiles in the original SIMBA simulation to the corresponding $N$-body simulation across the radial range $0.1 < r/r_{200} < 5$, concluding that the presence of baryons results in a decrease in dark matter density at the centers and intermediate regions of haloes owing to feedback. While our results for CAMELS-SIMBA are in good agreement across the same range of radii, in our work we have explored a wider range $0.01 < r/r_{200} < 10$, which shows that the dark matter density in fact greatly increases at small radial distances ($r/r_{200} < 0.04$). We note here that the innermost radial bin is spatially resolved for $M_{\rm halo} \gtrsim 10^{11.5}\,\rm{M_{\odot}}$, but is indeed approaching the resolution limit ($\sim$$1$\,kpc). Our results are also in qualitative agreement with \cite{schaller_2015_eagle_analysis} and \cite{Sorini_2024_BR} showing that including baryons increases the innermost halo densities in the EAGLE and IllustrisTNG simulations, respectively. 
For the halo mass range analyzed ($10^{11}\,{\rm M}_{\odot} < M_{\rm halo} < 10^{14}\,\rm{M_{\odot}}$), central dark matter densities also generally increase in the presence of baryons, as expected from traditional adiabatic contraction scenarios \citep{Blumenthal1986,Gnedin_2004_contraction}.
For enclosed mass profiles, we find similar results to the simulations including AGN feedback in \cite{Velliscig2014}. In particular, Swift-EAGLE shows the most similarities across different halo mass ranges, though this is not unexpected given the common lineage of OWLS and EAGLE. At lower halo masses, SIMBA, IllustrisTNG, and ASTRID show weaker effects relative to $N$-body, likely due to differences in stellar feedback implementation compared to Swift-EAGLE.
However, we note that given the limited resolution of CAMELS ($\sim$1\,kpc), we cannot address the impact of stellar feedback on the mass profiles of dwarf galaxies with $M_{\rm halo} \sim 10^{10}\,\rm{M_{\odot}}$. Previous works have identified this mass scale to be greatly impacted by supernova feedback  \citep[e.g.,][]{Governato2012,Chan_2015_FIRE_BR,Weinberg2015,Lazar2020,Sales2022}, and higher-resolution simulations exploring cosmological and astrophysical parameter variations are crucial to understand the nature of dark matter \citep{Rose2025}.

\begin{figure*}
	\includegraphics[width=\textwidth]{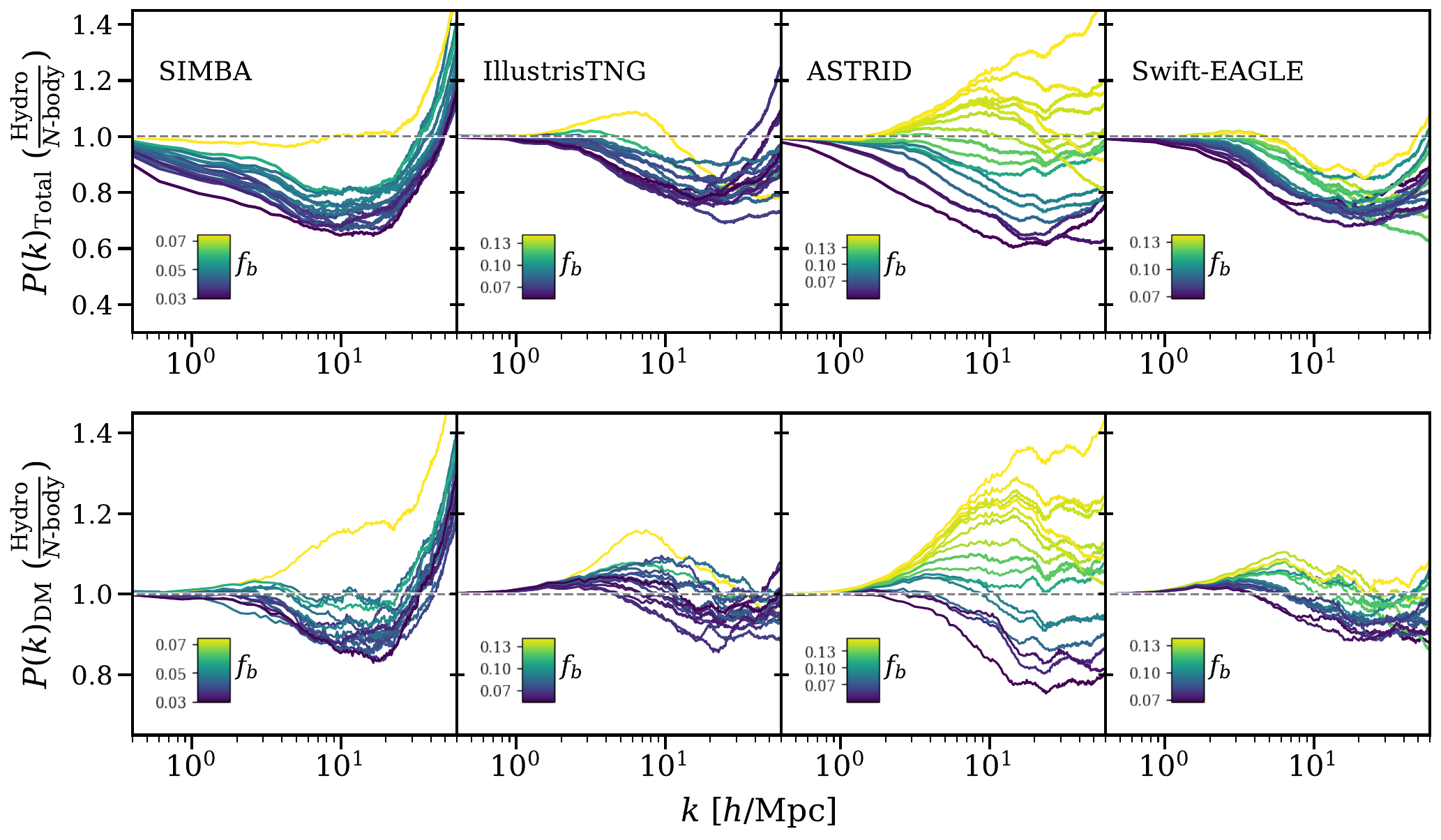}
    \vspace{-0.2in}
    \caption{Ratio of total matter (top row) and dark matter (bottom row) power spectra at $z=0$ from the hydrodynamic to $N$-body simulations in the six-parameter 1P sets of SIMBA, IllustrisTNG, ASTRID, and Swift-EAGLE that vary feedback parameters. Color scale corresponds to the average baryon fraction of the 10 most massive haloes in each hydrodynamic simulation. There exists a clear correlation between the baryon fraction in massive haloes and impact of baryons on the matter power spectrum.}
    \label{fig:pk_fb}
\end{figure*}

The impact of baryons on the total matter power spectrum is of great interest in light of current and upcoming weak lensing surveys that will be able to measure the total matter distribution with unprecedented precision \citep{Chisari_2019}. \cite{van_Daalen_2011} and later \cite{van_Daalen_2020} investigated the baryonic impact on the matter power spectrum in the OWLS and BAHAMAS simulations. Without AGN feedback, the suppression of power (relative to $N$-body) due to baryons was $\sim$$1\%$ for $0.8 \lesssim k \lesssim 6\,h\,\rm{Mpc^{-1}}$, while the power increased by $\sim$$10\%$ for $k \sim 10\,h\,\rm{Mpc^{-1}}$. Including AGN feedback suppressed power by several tens of percent for $1 \lesssim k \lesssim 10\,h\,\rm{Mpc^{-1}}$. We find qualitatively similar results, particularly when varying AGN feedback parameters (Figure \ref{fig:pk_variations_tot}): Simulations with more efficient AGN feedback show a greater suppression of the total matter power spectrum. Interestingly, as discussed in \cite{Gebhardt_2024a}, increasing the $A_{\rm SN2}$ parameter (mass loading factor) in SIMBA results in a smaller impact from baryons due to the non-linear coupling of stellar and AGN feedback. A similar effect may be occurring in IllustrisTNG, ASTRID, and Swift-EAGLE when varying the $A_{\rm SN1}$ parameter: Overly efficient SNe feedback in galaxies suppressing the growth of SMBHs that would otherwise produce strong AGN feedback (see also \citealt{Tillman_2023b_lyalpha, Delgado_2023_powerspec, Medlock_2024_fb_energy}).

We also find qualitatively similar results for the back-reaction effect on dark matter compared to \citet{van_Daalen_2011, van_Daalen_2020}, who found that when including AGN feedback, the dark matter power spectrum can be suppressed as much as 10\% at $k \sim 10\,h\,\rm{Mpc^{-1}}$ depending on the simulation considered. We find that while the suppression of power in the fiducial IllustrisTNG and Swift-EAGLE simulations matches this level of suppression, the fiducial SIMBA simulations can exceed this value (owing to stronger AGN feedback) and the fiducial ASTRID simulations show instead very little impact. Much like for total matter, increasing the strength of AGN feedback increases the suppression of dark matter power, and increasing the strength of SNe feedback can lessen the effect (Figure \ref{fig:pk_variations}).

Recent studies have explored whether the impact of feedback on the matter power spectrum could alleviate the observed ``$S_8$ tension'' between predictions from observations of the cosmic microwave background (CMB) and present-day observations such as weak lensing \citep{Bigwood2025_DES}. 
Using simulations from the FLAMINGO project, \cite{McCarthy_2023_s8} found that baryonic effects alone are insufficient to fully resolve the tensions when comparing power and cross-power spectra of cosmic shear, CMB lensing, and the thermal Sunyaev-Zel'dovich (tSZ) effect. In a later analysis, \cite{McCarthy_2025_s8} compared FLAMINGO simulations to kSZ effect stacking measurements fit from galaxy-galaxy lensing measurements, finding that stronger feedback in simulations partially alleviates the difference in observed and predicted small-scale clustering. However, differences at large scales, inferred from tSZ power spectrum and its cross-spectrum with cosmic shear could not be resolved in this way. \cite{Schaller_2025_pk} showed that extreme feedback variations in an emulator built from FLAMINGO simulations (which produced a stronger baryonic impact on the matter power spectrum than the simulations analyzed here) can match weak lensing constraints, but conflicts with the cluster X-ray data used to constrain the fiducial simulation model (see also \citealt{Schaller_2025_analytic_Pk} for analytical predictions of baryonic effects on the matter power spectrum using FLAMINGO simulations). Our results here further highlight that the impact of baryons on the matter power spectrum widely varies depending on galaxy formation model and subgrid parameter choices, with the caveat that the wide parameter variations in CAMELS are not observationally constrained and thus do not necessarily represent the true uncertainty.

Our results are also in qualitative agreement with \cite{Wright_2024_baryoncycle}, who compared the role of stellar and AGN feedback for the baryon cycle in the original SIMBA, IllustrisTNG, and EAGLE simulations. Stellar feedback (dominating in haloes of $M_{\rm halo} < 10^{12}\,\rm{M_{\odot}}$) was found to be strong enough to eject a significant amount of baryonic mass from haloes in EAGLE, eject some mass in SIMBA, and only recycle gas within the halo in IllustrisTNG. AGN feedback (dominating in haloes of $M_{\rm halo} > 10^{12}\,\rm{M_{\odot}}$) was comparably ineffective at ejecting baryons out of haloes in EAGLE, while it was more efficient at doing so in IllustrisTNG and particularly strong in SIMBA. We can see similar trends in our results, including the halo baryon fractions (Figure \ref{fig:fb}). Indeed, in lower mass haloes, baryons are more easily ejected in SIMBA and Swift-EAGLE, resulting in lower halo masses relative to the counterpart $N$-body halo. Interestingly, this does not directly apply to the effect on dark matter mass, which is quite similar for SIMBA and IllustrisTNG. At larger masses, baryons are efficiently removed from haloes in both SIMBA and (to a lesser extent) IllustrisTNG, resulting in a decrease in both total and dark matter mass in haloes relative to $N$-body. These trends can also help interpret our results for the impact of baryonic physics on halo density profiles (Figure \ref{fig:total_den_pro}). At low masses, Swift-EAGLE and SIMBA (each with relatively strong stellar feedback) show a significant impact on halo density profiles relative to $N$-body simulations, while the impact is noticeably weaker in IllustrisTNG. At higher masses, however, SIMBA remains the strongest, while Swift-EAGLE (with comparably weaker AGN feedback) shows a lesser impact than IllustrisTNG. 

Besides highlighting key differences between SIMBA, IllustrisTNG, ASTRID, and Swift-EAGLE, the large suites of CAMELS parameter variations also enabled us to identify intrinsic dependencies of hydrodynamic to $N$-body halo mass ratios, radial profiles, and power spectra on cosmological parameters ($\Omega_{\rm m}$ and $\sigma_{8}$) and the systematic effects imprinted by the assumed stellar and AGN feedback efficiencies. 
Previously underappreciated, our results show that the impact of baryonic physics on the distribution of matter depends intrinsically on cosmology. Increasing  $\Omega_{\rm m}$ (at fixed $\Omega_{\rm b}$) in hydrodynamic simulations generally reduces the suppression of halo mass relative to $N$-body simulations, decreasing halo central densities, and reducing the suppression of the matter power spectrum. Importantly, the response to changes in cosmology depends on galaxy formation model, with variations of $\sigma_{8}$ producing contrasting effects depending on the feedback implementation.

As expected, changes in the strength of feedback can substantially impact halo masses, radial profiles, and power spectra. Previous works using CAMELS simulations have seen comparably wide ranges of results across subgrid models and feedback parameter variations.
\cite{Tillman_2023b_lyalpha} found that varying any of the four feedback parameters ($A_{\rm SN1}$, $A_{\rm SN2}$, $A_{\rm AGN1}$, or $A_{\rm AGN2}$) in SIMBA resulted in significant variation in the Ly$\alpha$ forest, concluding that while AGN feedback is the dominant method of injecting energy into the intergalactic medium, variations in stellar feedback can modulate the growth of SMBHs responsible for AGN feedback. We found similar results in \cite{Gebhardt_2024a} using CAMELS: owing to strong AGN feedback, SIMBA was able to directly redistribute $\sim$40\% of baryons (relative to their initial dark matter neighboring distribution) $>$1\,Mpc\,$h^{-1}$ away compared to $\sim$10\% for IllustrisTNG and ASTRID. Varying AGN feedback strength produces similar variation within the SIMBA model: when halving AGN jet speeds compared to the fiducial SIMBA model, 25\% of baryons spread farther than 1\,Mpc\,$h^{-1}$ compared to 55\% when AGN jet speeds were increased to double the fiducial value. \cite{Medlock_2024_fb_energy} used CAMELS to quantify the energy injected by feedback in the SIMBA and IllustrisTNG suites, finding that IllustrisTNG generally shows greater enclosed feedback energy in haloes, while the ``closure radius'' (see \citealt{Ayromlou_2022_closureradius}) was much larger for SIMBA, indicating that energy is injected and has a stronger effect farther away from haloes. 
Increasing the efficiency of stellar feedback in IllustrisTNG strongly reduced the effect of AGN feedback (by inhibiting SMBH growth), while variations in AGN feedback strength directly did not significantly affect the energy injected, as seen in previous works \citep[e.g.,][]{Busillo2023, Ni_2023_CAMELSastrid, Tillman_2023b_lyalpha,Medlock_2024_fb_energy}. Meanwhile, in SIMBA, varying AGN feedback parameters had a much more dramatic effect. Clearly, the interplay between SNe and AGN feedback is complex and depends on the subgrid model implementation, as we have demonstrated based on their combined effects on halo masses, radial profiles, and matter power spectra.

CAMELS also provides avenues to find mappings between physical quantities that are significantly affected by feedback. For example, \cite{Delgado_2023_powerspec} found significant differences in both the total matter power spectrum and in mean halo baryon fractions when varying subgrid feedback parameters for the SIMBA and IllustrisTNG models. Following the work of \cite{van_Daalen_2020}, they found a mapping between the baryonic suppression of the total matter power spectrum and halo baryon fractions. In Figure \ref{fig:pk_fb}, we explicitly show this connection by color-coding lines from Figures \ref{fig:cv_pk_total} and \ref{fig:cv_pk_br} based on the baryon fraction in the ten most massive haloes in the hydrodynamic simulation. For all models, simulations in which massive haloes have lower average baryon fractions show systematically stronger suppression of both total and dark matter power spectra. This result is in agreement with \cite{Salcido_2023_pkfb}, who found a similar correlation using the ANTILLES suite of simulations and developed a model to predict the baryonic impact on the matter power spectrum using only the mean baryon fraction of haloes. In \cite{Gebhardt_2024a}, we similarly found a clear connection between baryonic suppression of power and the median spreading of baryons relative to dark matter. Also using CAMELS simulations, \cite{Medlock_2025_frb} found a correlation between the spreading of baryons and fast radio bursts, enabling further observational constraints on the impact of baryons on the cosmic matter distribution.

CAMELS data is also designed to leverage machine learning techniques. For example, \cite{Pandey_2023_powerspec} used CAMELS data and tSZ effect measurements to train models capable of constraining the baryonic impact on the matter power spectrum to percent level precision. There are now also numerous additional works that have used CAMELS to effectively bypass the uncertainties of baryonic physics via machine learning tools that attempt to marginalize over these uncertainties while accurately predicting cosmological parameters \citep[e.g.][]{Villaescusa-Navarro_2020_marginalization_example, Villaescusa-Navarro_2021_marginalization_fields2, Villaescusa-Navarro_2021_marginalization_fields, Villanueva-Domingo_2022_marginalizing_cosmicgraphs, Shao_2022_marginalization_subhaloes, Perez_2022_camelsSAM, deSanti_2023_fieldlevelgalaxies}. 

Our results have emphasized the need for models that can explore the wide-ranging effects of baryonic physics, but cosmological hydrodynamic simulations are computationally expensive. There are a variety of methods to approximate the results of full cosmological hydrodynamic simulations, including baryonification methods \citep[e.g.][]{Schneider_2015_baryonification, Schneider_2019_baryonification, Weiss_2019_baryonification} and semi-analytical models \citep[e.g.][]{Kauffmann_1993_SAM, Somerville_1999_SAM, Croton_2006_SAM, Guo_2011_SAM}. Baryonification methods alter the 3D matter distribution in $N$-body simulations to match predictions (such as the total matter power spectrum) from hydrodynamic simulations and show great promise in extrapolating to large volumes representative of cosmological surveys (see e.g. \citealt{Schneider_2025}). However, these methods typically need to tune parameters to match existing physical models. 
Semi-analytic models (SAMs) can predict galaxy properties at a fraction of the cost of full hydrodynamic simulations through the use of dark matter halo merger trees and simplified flow equations \citep{Baugh_2006_SAMs, Somerville_2015}. While they are tuned to observations, SAMs too must parameterize many physical properties including feedback from SNe and AGN and do not model the 3D matter distribution of baryons \citep{Baugh_2006_SAMs, Somerville_2015}. A complementary strategy is data-driven emulation, where the impact of baryonic physics on the matter distribution, including dependence on cosmological and subgrid parameters, can be learned by generative machine learning models to accurately recreate the effects of specific galaxy formation models at an arbitrary choice of parameters. Baryonification via emulation, the topic of future work, is a promising avenue to recreate realistic matter distributions reflecting the impact of baryonic physics at the volumes required for cosmological analyses.


\section{Conclusions}
\label{Conclusions}

We have significantly expanded the scope of previous works exploring the impact of baryons on the cosmic matter distribution by using thousands of simulations encompassing a broad parameter space across four different galaxy formation models in CAMELS. Overall, the effect of baryons on the total matter distribution is quite significant, affecting many scales relevant to astrophysical and cosmological analyses. In this work, we have investigated the effects of baryons on the total and dark matter components of halo masses, matter density profiles, and matter power spectra. We summarize our key findings below:

\begin{itemize}

  \item Halo masses, radial profiles, and matter power spectra all show significant variation depending on the galaxy formation model, specific subgrid parameters, cosmological parameters, and redshift (though we have omitted redshift analysis for all but halo masses).

  \item In general, the inclusion of baryons affects the dark matter distribution in a similar, though less extreme, manner as the total matter distribution. Correcting for this back-reaction effect is crucial to accurately model the distribution of dark matter and total matter.

  \item The masses of haloes in hydrodynamic simulations are generally reduced relative to corresponding $N$-body simulations owing to feedback ejecting or preventing the infall of gas (Figure~\ref{fig:total_halo_mass}). This effect generally correlates with halo baryon fraction and halo mass range, and can vary strongly across different simulations (Figure~\ref{fig:fb}). For example, halo masses in SIMBA are reduced by as much as 30\% in the lower mass range ($M_{\rm halo} \sim$10$^{10}$$\, {\rm M_{\odot}}$) at $z=0$, while halo masses in ASTRID are reduced by only 15\% in this range. 
  
  \item The dark matter mass in haloes is also generally reduced in hydrodynamic simulations and can see a similar level of variation across models (Figure~\ref{fig:dm_halo_mass}), with the dark matter masses reduced by as much as 20\% in SIMBA and only $\sim 5\%$ in ASTRID in the most extreme cases.

  \item By comparing density and enclosed mass profiles relative to $N$-body simulations (Figures~\ref{fig:total_den_pro} and~\ref{fig:total_mass_pro}), we find that matter in hydrodynamic haloes is more concentrated towards the center (due to radiative cooling) and reduced farther out (due to feedback processes). Back-reaction drives similar variation in dark matter profiles relative to $N$-body simulations. These effects are strongest in SIMBA and weakest in ASTRID, but the variance across models lessens for the most massive haloes.
  
  \item Both the total and dark matter power spectrum are strongly affected by baryons (Figures~\ref{fig:cv_pk_total} and~\ref{fig:cv_pk_br}). In the fiducial SIMBA model, power was suppressed relative to $N$-body simulations by $\sim$30\% and $\sim$15\% at $k \sim 10\,h\,\rm{Mpc^{-1}}$ for total and dark matter, respectively. IllustrisTNG and Swift-EAGLE both predict $\sim$25\% total matter power suppression and $\sim$$10\%$ dark matter power suppression at the same scale. ASTRID showed the weakest effects with $\sim$$15\%$ total matter power suppression and very minimal impact on the dark matter power spectrum.

  \item Our results depend intrinsically on cosmological and astrophysical parameters (Figures~\ref{fig:halo_mass_variations}, \ref{fig:total_mass_pro_variations}, \ref{fig:dm_mass_pro_variations}, \ref{fig:pk_variations_tot}, and~\ref{fig:pk_variations}). Increasing  $\Omega_{\rm m}$ (at fixed $\Omega_{\rm b}$) generally reduces the impact of baryonic physics relative to $N$-body simulations, with hydrodynamic simulations reducing the suppression of halo mass, decreasing halo central densities, and reducing the suppression of the matter power spectrum. The response to changes in cosmology depends on baryonic physics implementation, with variations of $\sigma_{8}$ producing contrasting effects depending on the galaxy formation model.
  
  \item Increasing the strength of SNe feedback decreases the central densities of haloes in most cases. Conversely, increasing SNe feedback strength can in many cases result in less suppression of the matter power spectrum owing to the negative impact on SMBH growth yielding a reduction of AGN feedback.
  
  \item Increasing the strength of AGN feedback generally decreases the masses of haloes in hydrodynamic simulations relative to $N$-body and results in greater suppression of the matter power spectrum.

  \item Similar to previous works, we find that the baryon fraction of massive haloes correlates with the impact of baryonic physics on the matter power spectrum (Figure~\ref{fig:pk_fb}).  However, the baryon fraction--power spectrum connection depends on the details of the galaxy formation model implementation.
\end{itemize}

Understanding how baryonic processes affect the surrounding matter distribution lies at the crossroads of cosmology and galaxy formation physics. To most efficiently use the data from cosmological surveys, we need to learn the extent of the uncertainties these processes can produce. In this work, we aimed to characterize these uncertainties for observationally relevant quantities in four state-of-the-art cosmological hydrodynamic simulations. Like many recent works, our results have again demonstrated the significant differences across galaxy formation models. Similarly, in exploring the wide parameter space in CAMELS, we have shown the extent to which cosmological and subgrid parameters can affect our quantities of interest, including the significant impact of feedback parameter variations on the dark matter distribution. Unfortunately, the ability to compare different models over a broad space of parameters comes with a cost.
Due to the relatively low resolution and volumes in CAMELS, it was not feasible to explore in detail the back-reaction effect in the low mass ($M_{\rm halo}$ < $10^{11}\,\rm{M_{\odot}}$) and high mass ($M_{\rm halo}$ > $10^{14}\,\rm{M_{\odot}}$) regimes at the level of individual halo radial profiles, which naturally limits the scope of this analysis. Additionally, the smaller CAMELS volumes limit the number of modes available in power spectra analysis, including observationally relevant scales (see \citealt{Schaller_2025_pk}). 
Exploring group-scale haloes ($M_{\rm halo}$ > $10^{14}\,\rm{M_{\odot}}$) across cosmological and astrophysical parameter variations is already possible  using dedicated zoom-in simulations targeting massive haloes with the IllustrisTNG model (\citealt{Lee_2024_carpool}), and second-generation CAMELS simulations using $(\text{50\,Mpc}\,h^{-1})^{3}$ volumes (Genel et al., {\it in prep.}) will alleviate these limitations.

\section*{Acknowledgements}
We thank Leah Bigwood for sharing the cosmic shear and kSZ total matter power spectra constraints used in Figure \ref{fig:cv_pk_total}. 
We thank the referee for constructive comments that helped improve the paper.
The CAMELS simulations were performed on the supercomputing facilities of the Flatiron Institute, which is supported by the Simons Foundation. DAA acknowledges support from NSF grant AST-2108944 and CAREER award AST-2442788, NASA grant ATP23-0156, STScI JWST grants GO-01712.009-A, AR-04357.001-A, and AR-05366.005-A, an Alfred P. Sloan Research Fellowship, and Cottrell Scholar Award CS-CSA-2023-028 by the Research Corporation for Science Advancement. 
D.N. and I.M. acknowledge support by the NSF grant AST-2511137.
\section*{Data Availability}
The CAMELS simulations are publicly available (see \citealt{CAMELS_data_release} and \citealt{Ni_2023_CAMELSastrid}). More information and instructions for downloading the data can be found at https://camels.readthedocs.io.



\bibliographystyle{mnras}
\bibliography{mattsbib} 


\appendix
\section{Effect of parameter variations on halo density profiles}
\label{appendix1}
\begin{figure*}
	\includegraphics[width=\textwidth]{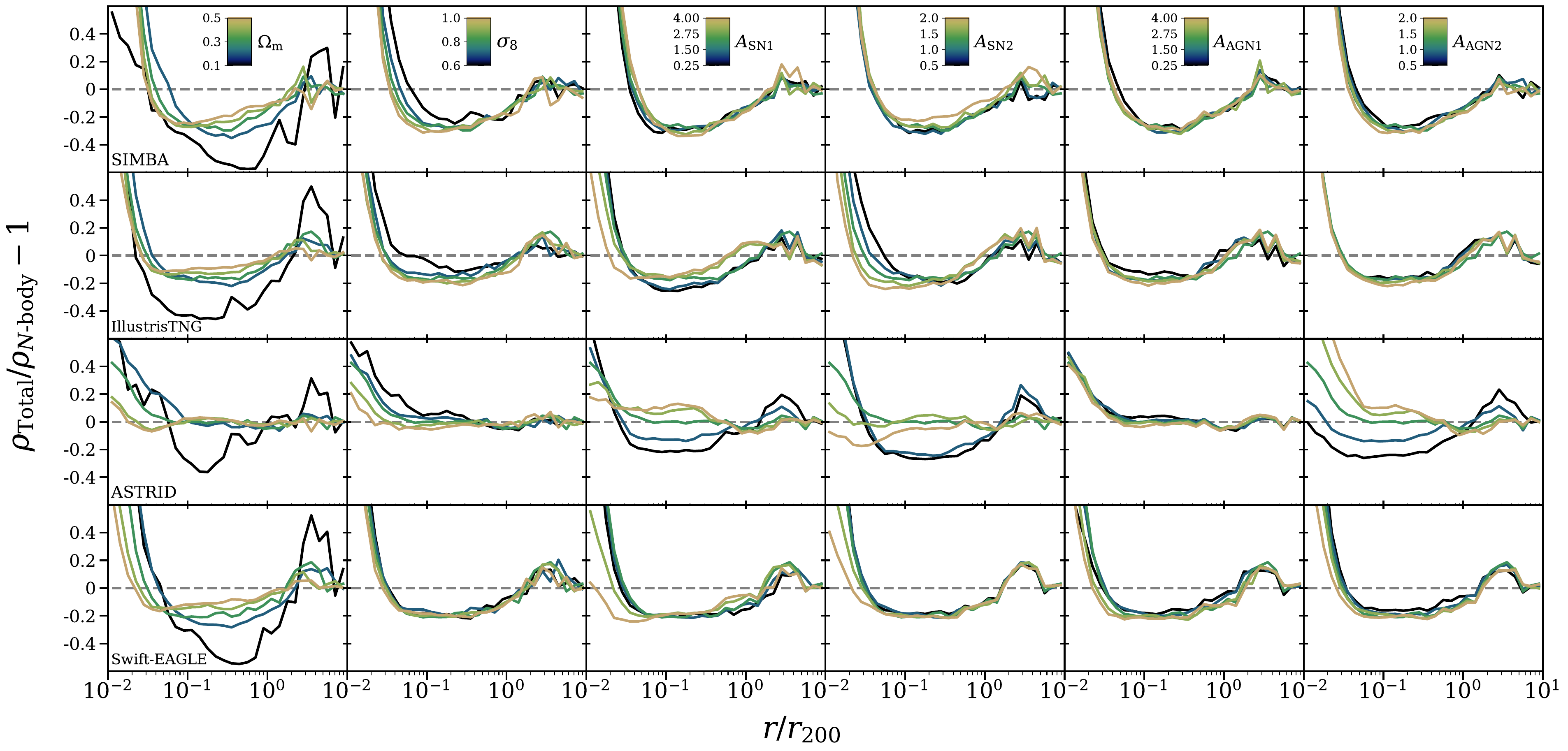}
    \vspace{-0.2in}
    \caption{Relative difference between hydrodynamic and $N$-body total matter density profiles at $z=0$ for all matched haloes with $M_{\rm halo} > 10^{12}\,\rm{M_{\odot}}$ in the six-parameter 1P set for SIMBA, IllustrisTNG, ASTRID, and Swift-EAGLE (same as the solid lines in the bottom panels of Figure \ref{fig:total_den_pro} but now for cosmological and feedback parameter variations). Each row corresponds to a different simulation suite, and each column corresponds to variations of a different parameter, as noted in the first row and column. Lines of different colors in each panel indicate the relative difference between density profiles for simulations varying only the corresponding parameter, as indicated by the color bar. As with halo enclosed mass profiles, halo density profiles depend strongly on cosmology at fixed baryonic physics, and the impact of baryonic physics is very sensitive to galaxy formation model and subgrid parameters.}
    \label{fig:den_pro_variations}
\end{figure*}

Figure \ref{fig:den_pro_variations} shows the relative difference between hydrodynamic and $N$-body total matter density profiles at $z=0$ for all matched haloes with $M_{\rm halo} > 10^{12}\,\rm{M_{\odot}}$ (same as the solid lines in the bottom panels of Figure \ref{fig:total_den_pro} but now for cosmological and feedback parameter variations). 
SNe feedback parameter variations significantly alter the mass distribution within haloes (generally decreasing central densities with increased SNe efficiency), while variations in AGN feedback parameters have generally weaker effects (the exception being $A_{\rm{AGN2}}$ in ASTRID--energy per unit accretion in thermal mode--which can strongly increase the central densities of haloes). 

\begin{figure*}
	\includegraphics[width=\textwidth]{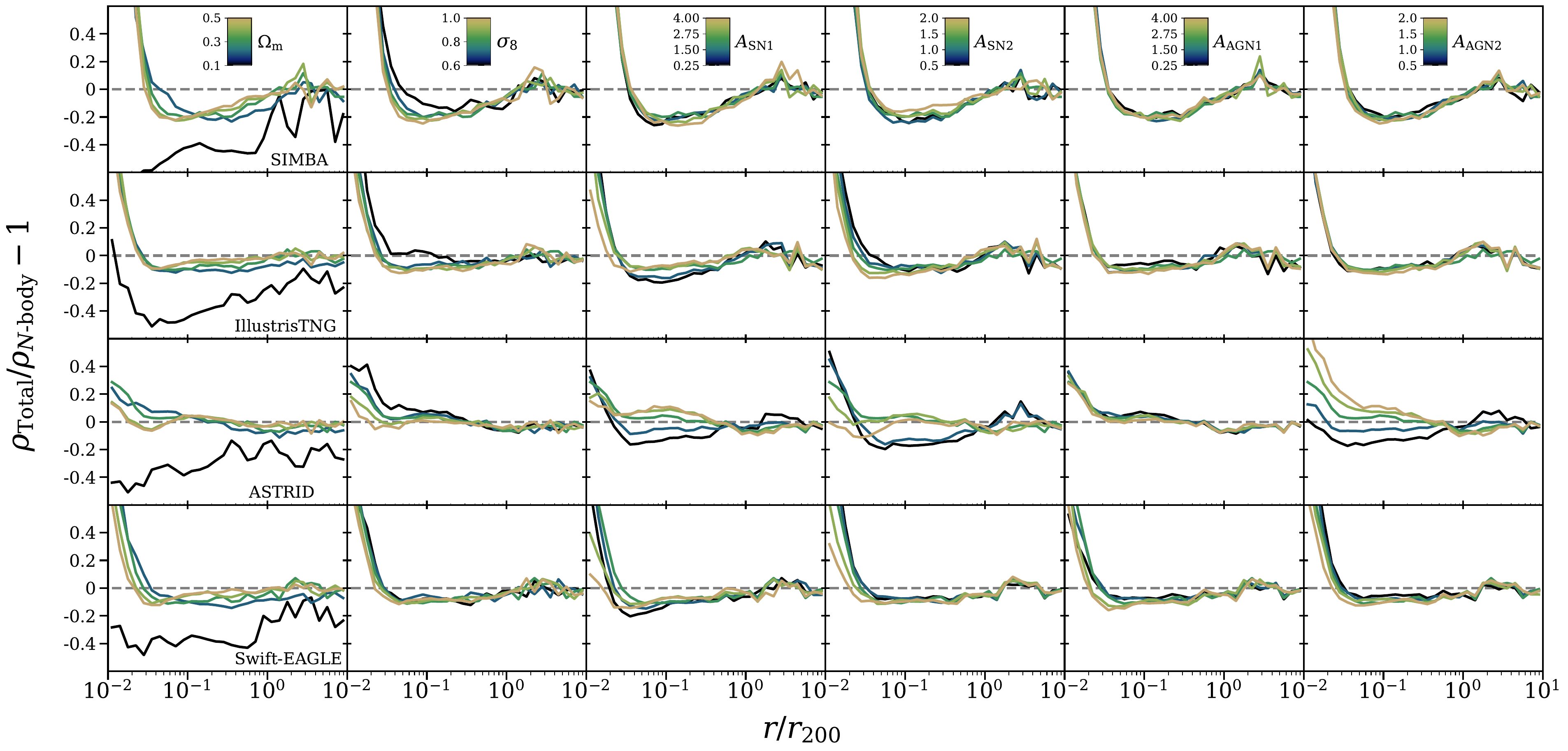}
    \vspace{-0.2in}
    \caption{Relative difference between hydrodynamic and $N$-body dark matter density profiles at $z=0$ for all matched haloes with $M_{\rm halo} > 10^{12}\,\rm{M_{\odot}}$ in the six-parameter 1P set for SIMBA, IllustrisTNG, ASTRID, and Swift-EAGLE (same as the dashed lines in the bottom panels of Figure \ref{fig:total_den_pro} but now for cosmological and feedback parameter variations). Dark matter particle masses are scaled up to be the same as in the $N$-body runs when calculating the hydrodynamic density profiles such that the relative difference would be zero at all radii in the absence of back-reaction effects. Each row corresponds to a different simulation suite, and each column corresponds to variations of a different parameter, as noted in the first row and column. Lines of different colors in each panel indicate the relative difference between density profiles for simulations varying only the corresponding parameter, as indicated by the color bar. Variations in dark matter density profiles generally follow the impact of cosmological and feedback parameters seen for total matter density profiles (Figure \ref{fig:den_pro_variations}), showing substantial back-reaction effects.}
    \label{fig:dm_den_pro_variations}
\end{figure*}

Figure \ref{fig:dm_den_pro_variations} is similar to Figure \ref{fig:den_pro_variations} but for dark matter density profiles. As with halo masses and power spectra, dark matter density profiles closely follow the trends of total matter with slightly less deviation from $N$-body simulations. Nevertheless, the back-reaction effects seen in halo density profiles are significant.

\bsp	
\label{lastpage}
\end{document}